\documentclass{IEEEtran}
\usepackage{cite}
\usepackage{amsmath,amssymb,amsfonts}
\usepackage{algorithmic}
\usepackage{graphicx}
\usepackage{textcomp}
\usepackage{xcolor}
\usepackage{bm}

\usepackage{amsfonts,amssymb,amsbsy,amsmath}
\usepackage{svg}
\usepackage{multirow}
\usepackage{float}
\usepackage{threeparttable}
\usepackage{import}
\usepackage{multicol}
\usepackage{booktabs}

\usepackage{textcomp}
\usepackage{stfloats}
\usepackage{url}
\usepackage{verbatim}
\usepackage{graphicx}
\usepackage{cite}
\usepackage{hyperref}
\usepackage{amsfonts,amssymb,amsbsy,amsmath}
\usepackage{algorithm}
\usepackage{array}

\def\BibTeX{{\rm B\kern-.05em{\sc i\kern-.025em b}\kern-.08em
    T\kern-.1667em\lower.7ex\hbox{E}\kern-.125emX}}
\begin{document}
\title{Fully Scalable Polarization-Reconfigurable S/X-Band Shared-Aperture Phased Array for Ultra-Low Axial-Ratio Scanning}

\author{Mohamed Räsänen, Juha Ala-Laurinaho, Samuel de Jésus Ndimubandi, Eugenio Cano Muñoz, Xiaoliang Sun Wang,  Alfonso Tomás Muriel-Barrado, Andrea Di Giovanni, Raffaele Di Bari, Marco Alessandrini, José Manuel Fernández González, \IEEEmembership{Senior Member, IEEE}, and Ville Viikari, \IEEEmembership{Senior Member, IEEE}

\IEEEmembership{}
\thanks{This work was supported by ESA through NGAAGS project (4000138412/22/NL/AS), Business Finland through RF ECO3 project (5665/31/2024), Research Council of Finland, Grant number 371367, and Advancing Radio Technologies for the Evolution towards 6G (UPM-ARTE-6G, PID2024-157242OB-C41),  funded by MICIU/AEI/10.13039/501100011033 and FEDER, EU. The work of Mohamed Räsänen was supported in part by the Helsingin Puhelinyhdistys (HPY) Research Foundation. \it{(Corresponding author: Mohamed Räsänen.)}}
\thanks{Mohamed Räsänen, Juha Ala-Laurinaho, and Ville Viikari are with the Department of Electronics and Nanoengineering, Aalto University, 00076 Espoo, Finland (e-mail: firstname.lastname@aalto.fi).}
\thanks{S.J. Ndimubandi, E.C. Muñoz, X.S. Wang, and J.M.F. González are with the Information Processing and Telecommunications Center, E.T.S.I. Telecomunicación, Universidad Politécnica de Madrid (UPM), Madrid, Spain (e-mail: josemanuel.fernandez.gonzalez@upm.es), S.J Ndimubandi is also affiliated with The Institute of Applied Physics, University of Electronic Science and Technology of China, Chengdu 610054, China.}
\thanks{Alfonso Tomás Muriel-Barrado is with the Escuela Politécnica Superior, Universidad Autónoma de Madrid, Madrid, Spain (e-mail: alfonsot.muriel@uam.es), and was previously affiliated with E.T.S.I UPM.}
\thanks{Andrea Di Giovanni is with SENER Aeroespacial (e-mail: andrea.digiovanni@aeroespacial.sener).}
\thanks{Raffaele Di Bari (e-mail: raffaele.dibari@ext.esa.int) and Marco Alessandrini (e-mail: marco.alessandrini@esa.int) are with the European Space Agency (ESA).}}

\maketitle

\begin{abstract}
This paper presents a modular S-/X-band shared-aperture phased-array antenna (SAPAA) for satellite-communication ground-station reception. The proposed architecture uses a repeatable unit cell that supports independent S- and X-band operation within the same physical aperture and enables arbitrary aperture scaling. Dual-polarized radiators are combined with calibrated complex receive coefficients to synthesize linear polarization (LP), right-hand circular polarization (RHCP), and left-hand circular polarization (LHCP). The design burden of the electrically large shared aperture is reduced by using theoretical estimates for scan matching and inter-band isolation before full shared-aperture verification. Simulated and measured results demonstrate axial ratios below 0.1\,dB in the target S- and X-band receiving bands over \(\mathbf{\pm50^\circ}\) scan range. The prototypes are validated using two approaches: passive measurements, where the element responses are measured individually, and RF system-on-chip-based active measurements, where all available receive channels are measured simultaneously. The results confirm that the proposed SAPAA provides wide-angle scanning, very high polarization purity, and polarization-reconfigurable operation for multi-mission SATCOM ground terminals.
\end{abstract}

\newpage
\section{Introduction}
\label{sec:introduction}
\begin{figure}[!t]
\centering
\def\svgwidth{250pt}
%% Creator: Inkscape 1.4.3 (0d15f75, 2025-12-25), www.inkscape.org
%% Accompanies image file 'SAPAA_Concept_V2.pdf' (pdf, eps, ps)
\begingroup%
  \makeatletter%
  \providecommand\color[2][]{%
    \errmessage{(Inkscape) Color is used for the text in Inkscape, but the package 'color.sty' is not loaded}%
    \renewcommand\color[2][]{}%
  }%
  \providecommand\transparent[1]{%
    \errmessage{(Inkscape) Transparency is used (non-zero) for the text in Inkscape, but the package 'transparent.sty' is not loaded}%
    \renewcommand\transparent[1]{}%
  }%
  \providecommand\rotatebox[2]{#2}%
  \newcommand*\fsize{\dimexpr\f@size pt\relax}%
  \newcommand*\lineheight[1]{\fontsize{\fsize}{#1\fsize}\selectfont}%
  \ifx\svgwidth\undefined%
    \setlength{\unitlength}{337.72149995bp}%
    \ifx\svgscale\undefined%
      \relax%
    \else%
      \setlength{\unitlength}{\unitlength * \real{\svgscale}}%
    \fi%
  \else%
    \setlength{\unitlength}{\svgwidth}%
  \fi%
  \global\let\svgwidth\undefined%
  \global\let\svgscale\undefined%
  \makeatother%
  \begin{picture}(1,0.67776688)%
    \lineheight{1}%
    \setlength\tabcolsep{0pt}%
    \put(0,0){\includegraphics[width=\unitlength,page=1]{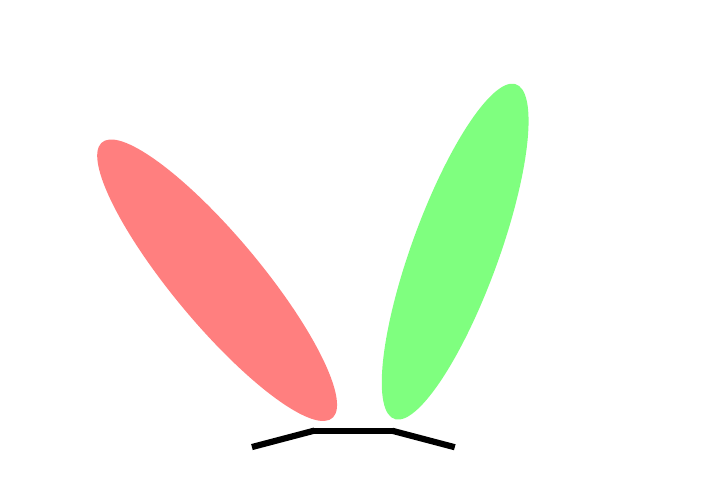}}%
    \put(0.42402579,0.11915297){\color[rgb]{0,0,0}\rotatebox{-50.468447}{\makebox(0,0)[rt]{\lineheight{1.25}\smash{\begin{tabular}[t]{r}S-band (2.2--2.29\,GHz)\end{tabular}}}}}%
    \put(0.59923467,0.16551816){\color[rgb]{0,0,0}\rotatebox{67.93205}{\makebox(0,0)[lt]{\lineheight{1.25}\smash{\begin{tabular}[t]{l}X-band (8--8.4\,GHz)\end{tabular}}}}}%
    \put(0.36400429,0.00605411){\color[rgb]{0,0,0}\makebox(0,0)[lt]{\lineheight{1.25}\smash{\begin{tabular}[t]{l}S/X-band SAPAA\end{tabular}}}}%
    \put(0,0){\includegraphics[width=\unitlength,page=2]{SAPAA_Concept_V2.pdf}}%
    \put(0.42686768,0.63330706){\color[rgb]{0.50196078,0.50196078,0.50196078}\rotatebox{10.974088}{\makebox(0,0)[t]{\lineheight{1.25}\smash{\begin{tabular}[t]{c}LEO\\CONSTELLATION\end{tabular}}}}}%
    \put(0,0){\includegraphics[width=\unitlength,page=3]{SAPAA_Concept_V2.pdf}}%
  \end{picture}%
\endgroup%

\caption{Illustration of the proposed S/X-band frustum shared-aperture phased array antenna concept for ground-station RX operation.}
\label{fig:SAPAA_concept}
\end{figure}

\IEEEPARstart{T}HE demand for high-capacity satellite-communication (SATCOM) links has accelerated the development of electronically steerable arrays that combine wide angular coverage, high gain, stable active matching, and polarization purity in compact apertures. Shared-aperture phased-array antennas (SAPAAs) are attractive for this purpose because multiple frequency bands can occupy the same physical aperture. The frustum-shaped S/X-band ground-station concept considered in this work is illustrated in Fig.~\ref{fig:SAPAA_concept}.

Shared-aperture arrays have been reported for Ku-/Ka-, K-/Ka-, X-/Ka-, and S-/C-band systems, demonstrating high isolation, wide bandwidth, and wide-angle scanning~\cite{Ding2019KuKaSharedAperture,Tong2022DualWidebandReflectarrays,Ding2022DualBand2DPhasedArray,Chang2022NearFieldCircularPolarizer,Hao2023KKaWideCoverageLEO,Zhang2023PlanarSimultaneousTxRxLEO,Liu2024LowProfileDualBandCP,Zhang2024WideAngleScanningCP,Xiao2025SCBandDualCP,Qing2025ModularKKaLTCC,Ji2025XKaCircularlyPolarized,Cai2025CoaxiallyLocatedKKaDualCP,Li2025KKaGroundTerminals,Mao2026CompactKKaDualCP}. However, most reported SAPAAs are restricted to fixed LP, fixed CP, or a predetermined pair of orthogonal polarization states. Their scanned AR values also remain several decibels in most cases, since embedded-element polarization imbalance, mutual coupling, finite-array effects, and calibration errors all perturb the synthesized CP state.

Array-level excitation control has been used previously to improve scanned circular-polarization performance in dual-polarized phased arrays \cite{Kim2022WidebandCPWideARScanning} and in shared-aperture CP arrays \cite{Liang2025KKaARSFSharedAperture}. However, prior demonstrations have generally targeted AR levels on the order of several decibels, or have been limited to fixed CP states. In contrast, the present work combines a shared S-/X-band aperture with calibrated receive-array circular-polarization synthesis to achieve below-0.1-dB AR over the target scan region while retaining LP/RHCP/LHCP reconfigurability. 

For ground-station SATCOM reception, ultra-low-AR CP synthesis provides higher polarization purity than the conventional 3-dB AR criterion. This is especially useful in RHCP/LHCP polarization-reuse links, where opposite-CP leakage contributes to co-channel interference. Ideally, reducing the AR from 3\,dB to 0.1\,dB improves opposite-CP isolation by approximately 30\,dB, which can help preserve the link quality required for high-throughput downlinking during the limited satellite pass time.

The proposed radiating aperture is an S-/X-band SAPAA unit cell intended for both planar apertures and frustum-shaped ground-station arrays. Since each face can scan over a wide angular sector, only a limited number of planar faces is required to cover the desired field of view, while the repeatable cell permits the aperture size to be scaled according to the required gain and G/T. The main novelties are:

\begin{enumerate}
    \item A scalable S-/X-band shared-aperture unit cell that combines a full-cell S-band radiator with a co-located X-band subarray in a balun-free, mechanically simple architecture designed to support high scan gain and high polarization purity.
    \item An experimentally validated array polarization-synthesis approach that uses calibrated complex coefficients of the dual-polarized embedded responses to realize LP, RHCP, and LHCP polarization states with below-0.1-dB scanned AR over the target bands and scan region.
    \item A fast design methodology in which analytical estimates of active impedance matching and inter-band coupling guide the element design before full shared-aperture verification.
\end{enumerate}

To the authors' knowledge, a phased-array implementation has not been demonstrated for 0.1-dB AR scanning over a practical scan range with shared aperture configuration. The concept is validated through passive measurements, where embedded responses are measured sequentially and combined offline, and RF system-on-chip (RFSoC)-based active measurements, where all available receive channels are acquired simultaneously. The CAD models of the fabricated prototypes are provided as supplementary material~\cite{SupplementaryCAD}. The rest of this work is organized as follows: Chapter~\ref{ch:design} explains the design methodology and excitation scheme of this work, followed by experimental validation in Chapter~\ref{ch:experimental}. The results are then compared to the state-of-the-art in Chapter~\ref{ch:discussion}, and the work is finally concluded in Chapter~\ref{ch:conclusion}.

\section{Design and Operation Principle}\label{ch:design}

The proposed SAPAA targets S-/X-band ground-station reception with wide-angle beam steering, sufficient inter-band isolation, and polarization-reconfigurable operation. The passive radiators are required to provide suitable matching, scan coverage, and baseline 3-dB AR beamwidth. The final sub-0.1-dB AR performance is obtained after shared-aperture integration and calibrated receive excitation. The target specifications are summarized in Table~\ref{tab:specs}.

\begin{table}[!t]
    \centering
    \caption{Target specifications for the dual-band phased array}
    \label{tab:specs}
    \begin{threeparttable}
    \begin{tabular}{@{}lcc@{}}
        \toprule
        Target specification & S-band & X-band \\ \midrule
        Impedance bandwidth (GHz) & 2.2--2.29 & 8.025--8.4 \\
        Steering range & $\pm 50^{\circ}$ & $\pm 50^{\circ}$ \\
        Inter-band isolation & $>15~\text{dB}$ & $>15~\text{dB}$ \\
        Baseline dipole axial ratio\tnote{a} & $\leq3~\text{dB}$ & $\leq3~\text{dB}$ \\
        Final integrated-array axial ratio\tnote{a} & $\leq0.5~\text{dB}$ & $\leq0.5~\text{dB}$ \\
        \bottomrule
    \end{tabular}
    \begin{tablenotes}
        \footnotesize
        \item[a] AR requirements apply to RHCP and LHCP operation.
    \end{tablenotes}
    \end{threeparttable}
\end{table}

\begin{figure}[t!]
\centering
\def\svgwidth{250pt}
\import{./figs}{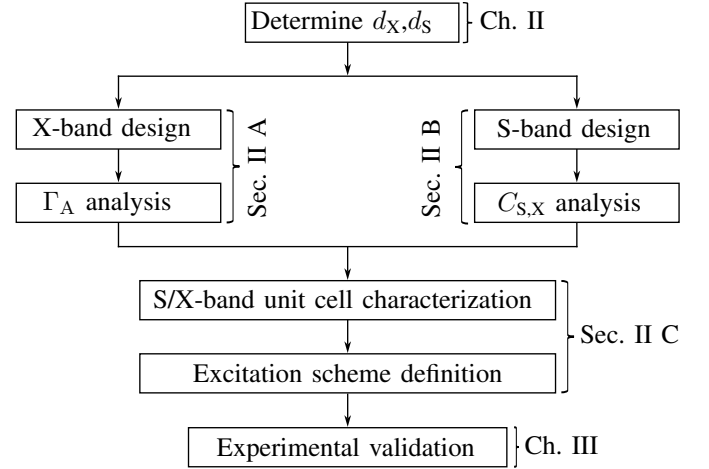}
\caption{Design and validation flowchart of the proposed SAPAA. \(d_\text{X}\), and \(d_\text{S}\) denote the band-respective lattice periods, \(\Gamma_\text{A}\) the active reflection coefficient, and \(C_\text{S,X}\) the inter-band coupling.}
\label{fig:design_flow}
\end{figure}

\begin{figure}[!t]
\centering
\def\svgwidth{250pt}
%% Creator: Inkscape 1.4.3 (0d15f75, 2025-12-25), www.inkscape.org
%% Accompanies image file 'SX_Design_Concept_V6.pdf' (pdf, eps, ps)
\begingroup%
  \makeatletter%
  \providecommand\color[2][]{%
    \errmessage{(Inkscape) Color is used for the text in Inkscape, but the package 'color.sty' is not loaded}%
    \renewcommand\color[2][]{}%
  }%
  \providecommand\transparent[1]{%
    \errmessage{(Inkscape) Transparency is used (non-zero) for the text in Inkscape, but the package 'transparent.sty' is not loaded}%
    \renewcommand\transparent[1]{}%
  }%
  \providecommand\rotatebox[2]{#2}%
  \newcommand*\fsize{\dimexpr\f@size pt\relax}%
  \newcommand*\lineheight[1]{\fontsize{\fsize}{#1\fsize}\selectfont}%
  \ifx\svgwidth\undefined%
    \setlength{\unitlength}{399.03044453bp}%
    \ifx\svgscale\undefined%
      \relax%
    \else%
      \setlength{\unitlength}{\unitlength * \real{\svgscale}}%
    \fi%
  \else%
    \setlength{\unitlength}{\svgwidth}%
  \fi%
  \global\let\svgwidth\undefined%
  \global\let\svgscale\undefined%
  \makeatother%
  \begin{picture}(1,0.59267496)%
    \lineheight{1}%
    \setlength\tabcolsep{0pt}%
    \put(0,0){\includegraphics[width=\unitlength,page=1]{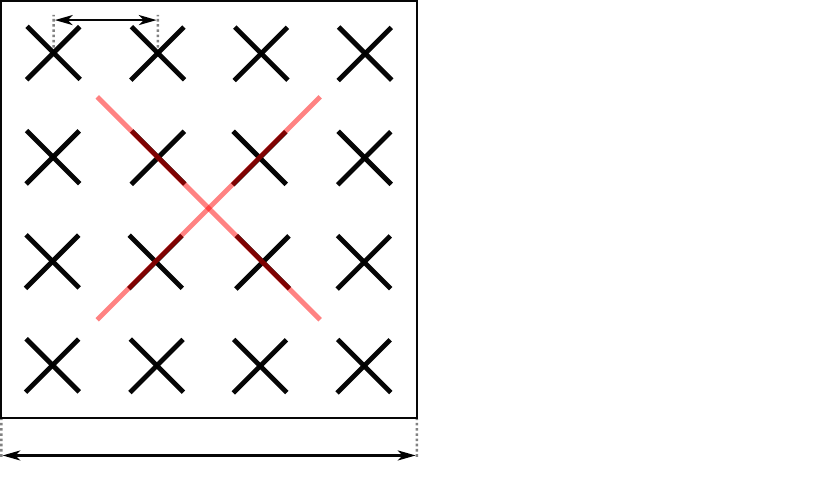}}%
    \put(0.12513893,0.52556965){\color[rgb]{0,0,0}\makebox(0,0)[t]{\lineheight{1.25}\smash{\begin{tabular}[t]{c}$d_\mathrm{X}$\end{tabular}}}}%
    \put(0.25405114,0.05744649){\color[rgb]{0,0,0}\makebox(0,0)[t]{\lineheight{1.25}\smash{\begin{tabular}[t]{c}$d_\mathrm{S}$\end{tabular}}}}%
    \put(0,0){\includegraphics[width=\unitlength,page=2]{SX_Design_Concept_V6.pdf}}%
    \put(0.24079818,0.00325976){\color[rgb]{0,0,0}\makebox(0,0)[t]{\lineheight{1.25}\smash{\begin{tabular}[t]{c}a)\end{tabular}}}}%
    \put(0.76925989,0.00325976){\color[rgb]{0,0,0}\makebox(0,0)[t]{\lineheight{1.25}\smash{\begin{tabular}[t]{c}b)\end{tabular}}}}%
  \end{picture}%
\endgroup%

\caption{a) Array configuration of the proposed SAPAA  lattice periods \(d_\mathrm{X}=17.5\,\text{mm}\) and \(d_\mathrm{S}=70\,\text{mm}\). b) Grating-lobe-free steering limits for the selected S-band lattice period.}
\label{fig:init_design}
\end{figure}

The design flow is shown in Fig.~\ref{fig:design_flow}. The lattice periods are selected from the grating-lobe condition at the upper edge of each band. The S-band period is chosen as \(d_\mathrm{S}=70~\mathrm{mm}\), corresponding to approximately \(0.53\lambda\) at 2.29\,GHz. As shown in Fig.~\ref{fig:init_design}, the grating-lobe boundary remains outside the specified \(\pm50^\circ\) scan range, with the first grating-lobe maximum approaching the visible region only near \(\theta_0\approx60^\circ\) in the principal planes. The \(4\times4\) X-band subarray inside one S-band cell gives \(d_\mathrm{X}=17.5\,\text{mm}\), or approximately \(0.49\lambda\) at 8.4\,GHz. 

Dipole radiators are utilized in both bands because they are simple, tunable, and directly compatible with dual-polarized receive beamforming. The X-band element is developed for scan-active matching, and the S-band element is shaped to provide both the required S-band embedded response and high X-band rejection. The electrically small X-band dipoles are omitted during the early S-band iterations. Their effect is monitored with the analytical coupling approximation in Section~II-B, and the final integrated unit cell is then verified using simulations with periodic boundary conditions.

The two bands are evaluated with metrics that reflect their different roles inside the shared cell. The S-band radiator occupies the full cell width and is assessed through embedded element patterns, whereas the X-band aperture is assessed through the scan gain envelope (SGE) and the axial ratio envelope (ARE) of the integrated \(4\times4\) subarray. The calibrated excitation scheme is then applied to synthesize LP, RHCP, and LHCP. Experimental validation uses a \(3\times4\) SAPAA prototype for shared-aperture and coupling characterization and a \(1\times8\) prototype for S-band beamforming.

\subsection{X-Band Element}\label{sec:XB}

\begin{figure}[t]
\centering
\def\svgwidth{250pt}
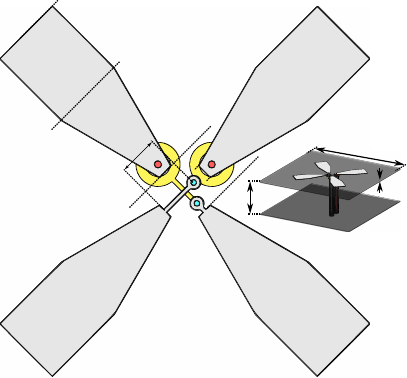
\caption{Initial X-band element unit cell with dimensions in mm. Dark gray denotes the top PCB layer, yellow denotes the bottom layer, red denotes the coaxial feed, and blue denotes vias. Dimensions highlighted in red are also used for the S-band element discussed in Section~\ref{sec:SB}.}
\label{fig:XB_UC}
\end{figure}

\begin{figure}[t]
\centering
\def\svgwidth{250pt}
%% Creator: Inkscape 1.4.3 (0d15f75, 2025-12-25), www.inkscape.org
%% Accompanies image file 'X_elem_initial_results_V4.pdf' (pdf, eps, ps)
\begingroup%
  \makeatletter%
  \providecommand\color[2][]{%
    \errmessage{(Inkscape) Color is used for the text in Inkscape, but the package 'color.sty' is not loaded}%
    \renewcommand\color[2][]{}%
  }%
  \providecommand\transparent[1]{%
    \errmessage{(Inkscape) Transparency is used (non-zero) for the text in Inkscape, but the package 'transparent.sty' is not loaded}%
    \renewcommand\transparent[1]{}%
  }%
  \providecommand\rotatebox[2]{#2}%
  \newcommand*\fsize{\dimexpr\f@size pt\relax}%
  \newcommand*\lineheight[1]{\fontsize{\fsize}{#1\fsize}\selectfont}%
  \ifx\svgwidth\undefined%
    \setlength{\unitlength}{716.22598002bp}%
    \ifx\svgscale\undefined%
      \relax%
    \else%
      \setlength{\unitlength}{\unitlength * \real{\svgscale}}%
    \fi%
  \else%
    \setlength{\unitlength}{\svgwidth}%
  \fi%
  \global\let\svgwidth\undefined%
  \global\let\svgscale\undefined%
  \makeatother%
  \begin{picture}(1,0.85653682)%
    \lineheight{1}%
    \setlength\tabcolsep{0pt}%
    \put(0,0){\includegraphics[width=\unitlength,page=1]{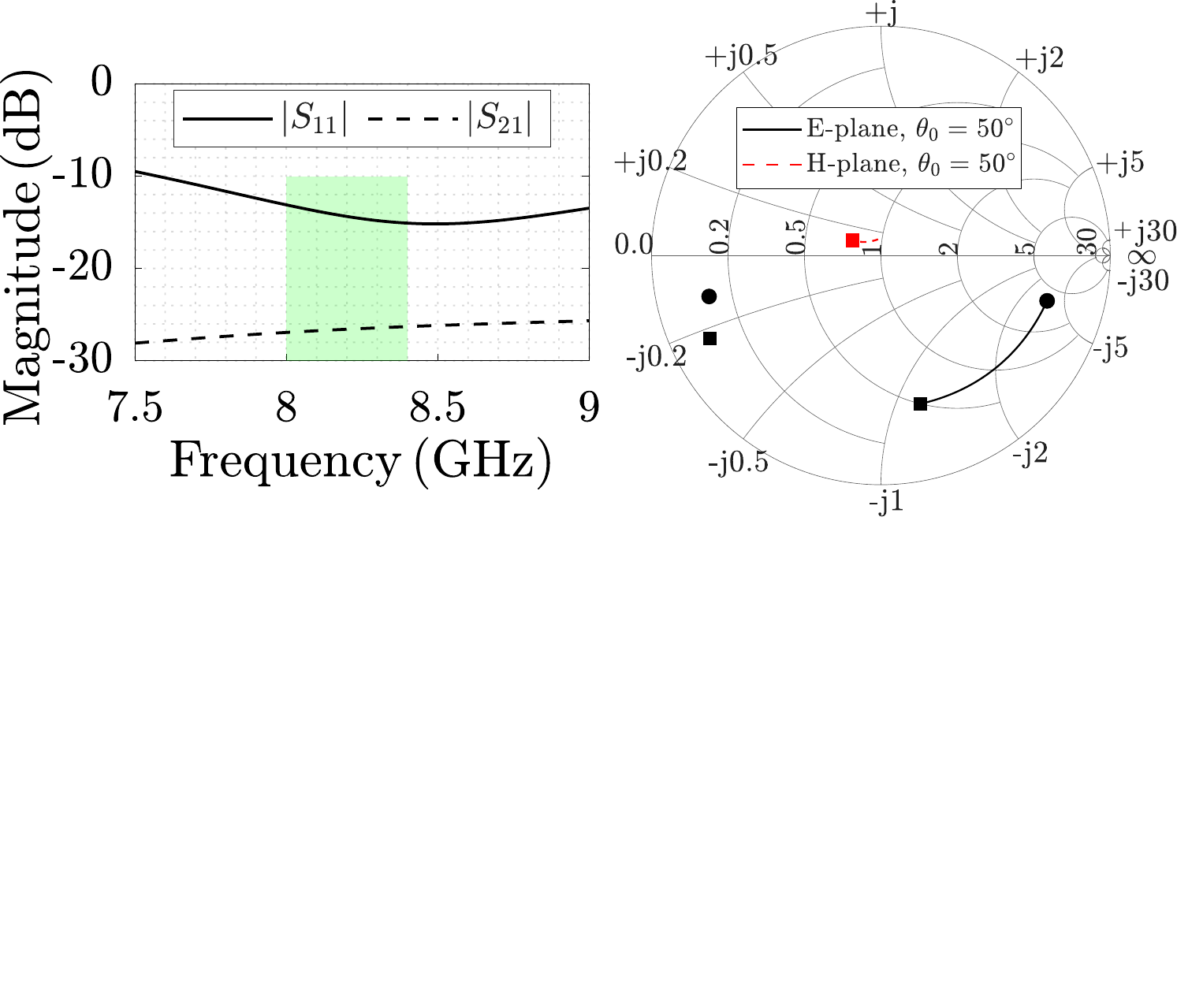}}%
    \put(0.61429972,0.59619844){\color[rgb]{0,0,0}\makebox(0,0)[lt]{\lineheight{1.25}\smash{\begin{tabular}[t]{l}\small{$f$=8\,GHz}\end{tabular}}}}%
    \put(0.61489677,0.56043659){\color[rgb]{0,0,0}\makebox(0,0)[lt]{\lineheight{1.25}\smash{\begin{tabular}[t]{l}\small{$f$=8.4\,GHz}\end{tabular}}}}%
    \put(0,0){\includegraphics[width=\unitlength,page=2]{X_elem_initial_results_V4.pdf}}%
    \put(0.15718782,0.30521174){\color[rgb]{0.50196078,0.50196078,0.50196078}\makebox(0,0)[lt]{\lineheight{1.25}\smash{\begin{tabular}[t]{l}$-3$\,dB\end{tabular}}}}%
    \put(0.2691012,0.3859752){\color[rgb]{0,0,0}\makebox(0,0)[lt]{\lineheight{1.25}\smash{\begin{tabular}[t]{l}a)\end{tabular}}}}%
    \put(0.74532032,0.3859752){\color[rgb]{0,0,0}\makebox(0,0)[lt]{\lineheight{1.25}\smash{\begin{tabular}[t]{l}b)\end{tabular}}}}%
    \put(0.53134819,0.0029237){\color[rgb]{0,0,0}\makebox(0,0)[lt]{\lineheight{1.25}\smash{\begin{tabular}[t]{l}c)\end{tabular}}}}%
  \end{picture}%
\endgroup%

\caption{Initial X-band unit cell performance: a) broadside active S-parameters, b) scan impedance for E- and H-plane scans at $\theta_0=50^\circ$, and c) E- and H-plane realized gain active embedded-element patterns, with $-3$-dB reference line with respect to the maximum amplitude.}
\label{fig:XB_init_results}
\end{figure}

The initial X-band unit cell is shown in Fig.~\ref{fig:XB_UC}a and is implemented on 0.254-mm-thick Rogers RO4350B \((\varepsilon_r=3.66,\tan\delta=0.0037)\). The broadside active S-parameters in Fig.~\ref{fig:XB_init_results}a show good matching and port isolation, aided by grounding the coaxial feeds to the reflector. At \(\theta_0=50^\circ\), the scan impedance in Fig.~\ref{fig:XB_init_results}b moves toward the capacitive region in the E-plane, which reduces the embedded-element gain in Fig.~\ref{fig:XB_init_results}c. Scan-dependent inductive compensation using pins is therefore introduced.

\begin{figure}[t]
\centering
\def\svgwidth{230pt}
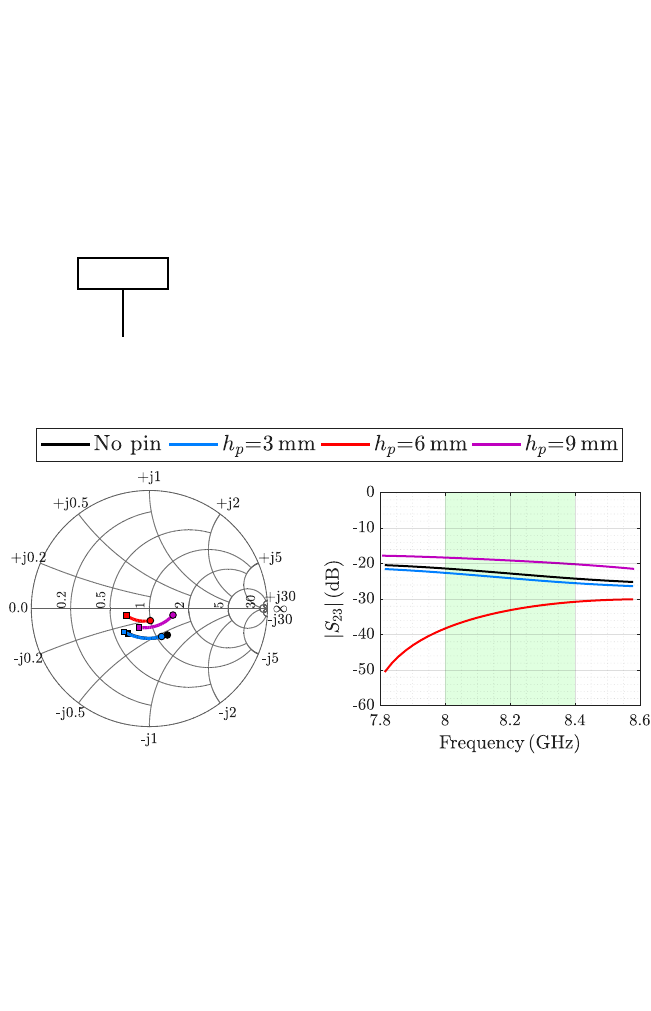
\caption{Impedance-compensation study using a simplified \(1\times3\) E-plane model: a) simulation setup, b) equivalent circuit interpretation, c) passive input reflection coefficient of port 2, d) coupling magnitude \(|S_{23}|=|S_{21}|\), and e) difference between the reflection phase of the center element and the coupling phase \(\angle S_{23}=\angle S_{21}\), where the green region denotes the out-of-phase region, and the red region the in-phase coupling region, respectively. The pin width is \(w_p=2\,\text{mm}\).}
\label{fig:XB_1x3}
\end{figure}

The compensation is studied using the simplified \(1\times3\) E-plane model in Fig.~\ref{fig:XB_1x3}a, whose equivalent circuit is shown in Fig.~\ref{fig:XB_1x3}b. The passive input reflection in Fig.~\ref{fig:XB_1x3}c is only weakly affected by the pins, indicating that the inductive loading does not strongly perturb the isolated element resonance. However, the coupling magnitude in Fig.~\ref{fig:XB_1x3}d and the reflection--coupling phase relation in Fig.~\ref{fig:XB_1x3}e change substantially. Among the studied heights \(h_p=3\), 6, and 9\,mm, the \(h_p=6\,\text{mm}\) pins, which reach the dipole level, provide high isolation, and therefore excellent active-matching behavior.

\begin{figure}[t]
\centering
\def\svgwidth{250pt}
%% Creator: Inkscape 1.4.3 (0d15f75, 2025-12-25), www.inkscape.org
%% Accompanies image file 'XB_1x3_Active_V3.pdf' (pdf, eps, ps)
\begingroup%
  \makeatletter%
  \providecommand\color[2][]{%
    \errmessage{(Inkscape) Color is used for the text in Inkscape, but the package 'color.sty' is not loaded}%
    \renewcommand\color[2][]{}%
  }%
  \providecommand\transparent[1]{%
    \errmessage{(Inkscape) Transparency is used (non-zero) for the text in Inkscape, but the package 'transparent.sty' is not loaded}%
    \renewcommand\transparent[1]{}%
  }%
  \providecommand\rotatebox[2]{#2}%
  \newcommand*\fsize{\dimexpr\f@size pt\relax}%
  \newcommand*\lineheight[1]{\fontsize{\fsize}{#1\fsize}\selectfont}%
  \ifx\svgwidth\undefined%
    \setlength{\unitlength}{528.87456939bp}%
    \ifx\svgscale\undefined%
      \relax%
    \else%
      \setlength{\unitlength}{\unitlength * \real{\svgscale}}%
    \fi%
  \else%
    \setlength{\unitlength}{\svgwidth}%
  \fi%
  \global\let\svgwidth\undefined%
  \global\let\svgscale\undefined%
  \makeatother%
  \begin{picture}(1,1.40073239)%
    \lineheight{1}%
    \setlength\tabcolsep{0pt}%
    \put(0,0){\includegraphics[width=\unitlength,page=1]{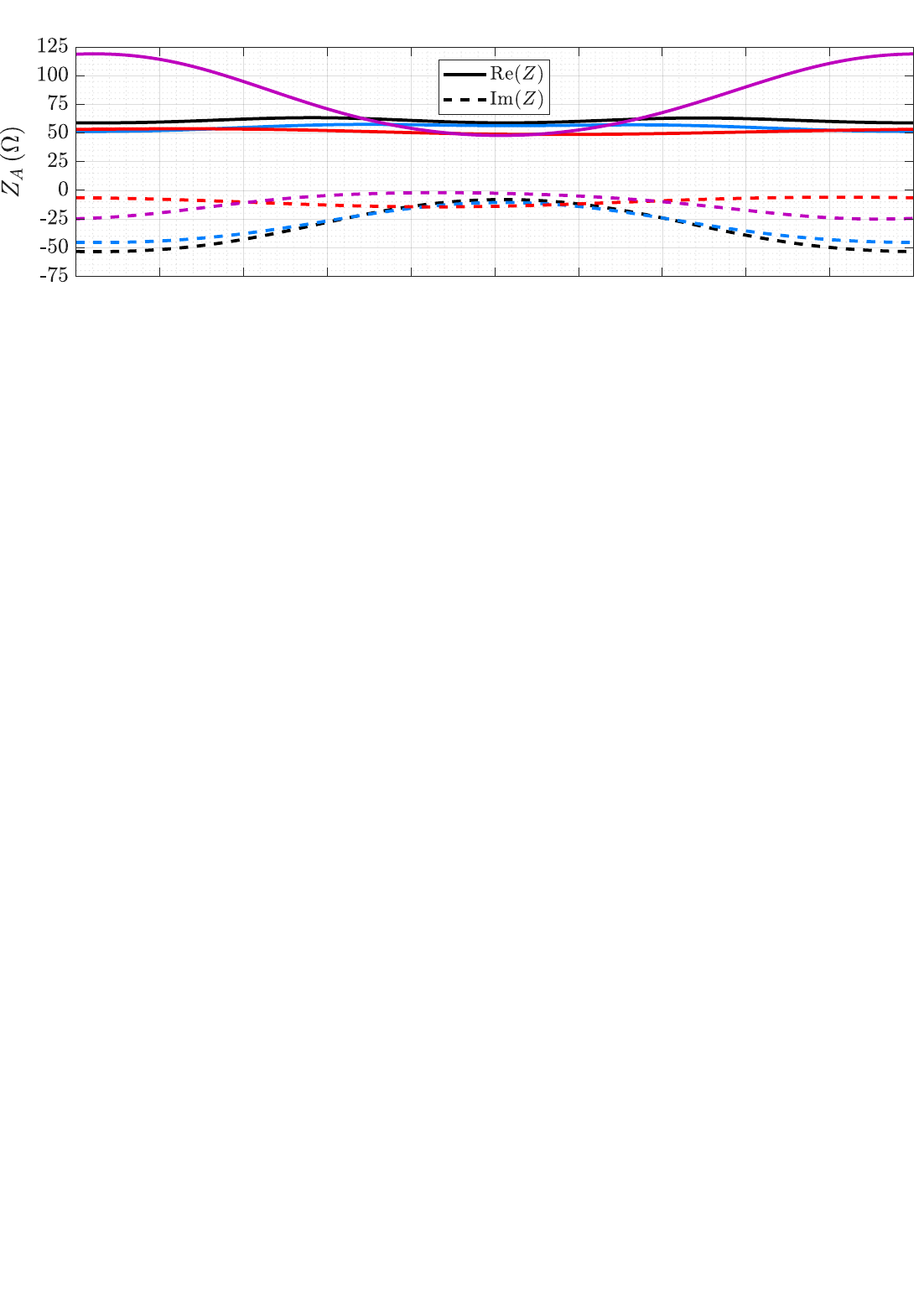}}%
    \put(0.53066257,1.06511724){\color[rgb]{0,0,0}\makebox(0,0)[lt]{\lineheight{1.25}\smash{\begin{tabular}[t]{l}a)\end{tabular}}}}%
    \put(0.53066257,0.75572118){\color[rgb]{0,0,0}\makebox(0,0)[lt]{\lineheight{1.25}\smash{\begin{tabular}[t]{l}b)\end{tabular}}}}%
    \put(0.53066257,0.38402109){\color[rgb]{0,0,0}\makebox(0,0)[lt]{\lineheight{1.25}\smash{\begin{tabular}[t]{l}c)\end{tabular}}}}%
    \put(0.53066257,0.00232314){\color[rgb]{0,0,0}\makebox(0,0)[lt]{\lineheight{1.25}\smash{\begin{tabular}[t]{l}d)\end{tabular}}}}%
    \put(0,0){\includegraphics[width=\unitlength,page=2]{XB_1x3_Active_V3.pdf}}%
  \end{picture}%
\endgroup%

\caption{Active impedance-compensation behavior of the simplified \(1\times3\) E-plane model: a) active input impedance, b) active-reactance contribution and theoretical approximation, c) active reflection coefficient across the target steering range at 8\,GHz, and d) active reflection coefficient at \(\theta_0=0^\circ\), \(30^\circ\), and \(50^\circ\) across frequency.}
\label{fig:XB_1x3_A}
\end{figure}

\begin{figure}[t!]
\centering
\def\svgwidth{250pt}
%% Creator: Inkscape 1.4.3 (0d15f75, 2025-12-25), www.inkscape.org
%% Accompanies image file 'X_5x5_comparison_V3.pdf' (pdf, eps, ps)
\begingroup%
  \makeatletter%
  \providecommand\color[2][]{%
    \errmessage{(Inkscape) Color is used for the text in Inkscape, but the package 'color.sty' is not loaded}%
    \renewcommand\color[2][]{}%
  }%
  \providecommand\transparent[1]{%
    \errmessage{(Inkscape) Transparency is used (non-zero) for the text in Inkscape, but the package 'transparent.sty' is not loaded}%
    \renewcommand\transparent[1]{}%
  }%
  \providecommand\rotatebox[2]{#2}%
  \newcommand*\fsize{\dimexpr\f@size pt\relax}%
  \newcommand*\lineheight[1]{\fontsize{\fsize}{#1\fsize}\selectfont}%
  \ifx\svgwidth\undefined%
    \setlength{\unitlength}{401.66197289bp}%
    \ifx\svgscale\undefined%
      \relax%
    \else%
      \setlength{\unitlength}{\unitlength * \real{\svgscale}}%
    \fi%
  \else%
    \setlength{\unitlength}{\svgwidth}%
  \fi%
  \global\let\svgwidth\undefined%
  \global\let\svgscale\undefined%
  \makeatother%
  \begin{picture}(1,0.9224427)%
    \lineheight{1}%
    \setlength\tabcolsep{0pt}%
    \put(0.49084016,0.47859792){\color[rgb]{0,0,0}\makebox(0,0)[lt]{\lineheight{1.25}\smash{\begin{tabular}[t]{l}a)\end{tabular}}}}%
    \put(0.49084016,0.00504757){\color[rgb]{0,0,0}\makebox(0,0)[lt]{\lineheight{1.25}\smash{\begin{tabular}[t]{l}b)\end{tabular}}}}%
    \put(0.94994018,0.90327082){\color[rgb]{0,0,0}\makebox(0,0)[rt]{\lineheight{1.25}\smash{\begin{tabular}[t]{r}Gain\,(dBic)\end{tabular}}}}%
    \put(0.95184091,0.47351458){\color[rgb]{0,0,0}\makebox(0,0)[rt]{\lineheight{1.25}\smash{\begin{tabular}[t]{r}AR\,(dB)\end{tabular}}}}%
    \put(0,0){\includegraphics[width=\unitlength,page=1]{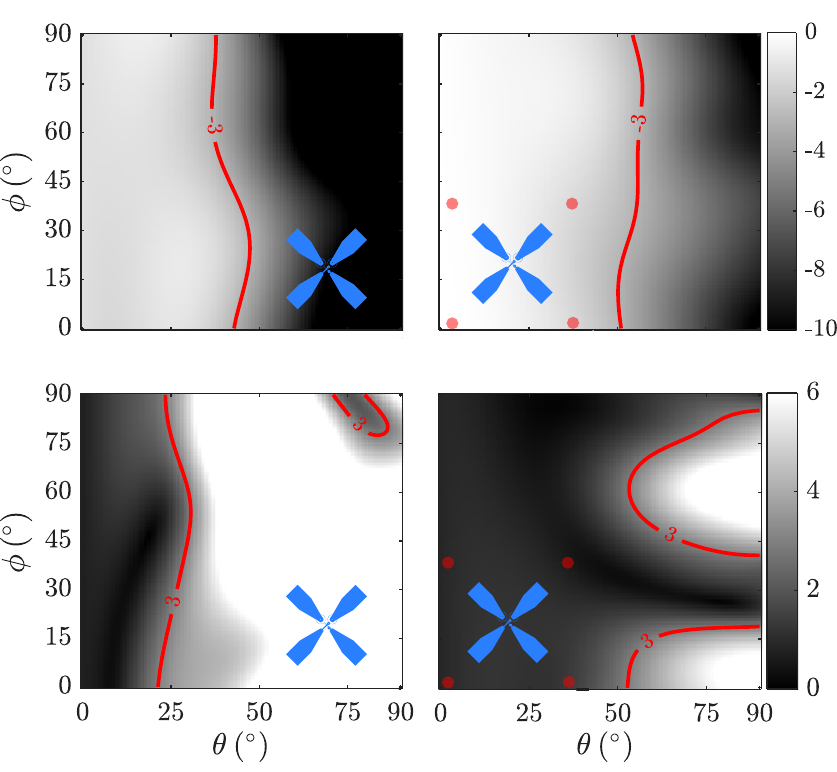}}%
  \end{picture}%
\endgroup%

\caption{Comparison of the original (left) and pin-enhanced (right) X-band dipole embedded-element performance in a \(5\times5\) array, a) circularly polarized embedded element gain and b) axial ratio.}
\label{fig:XB_5x5}
\end{figure}
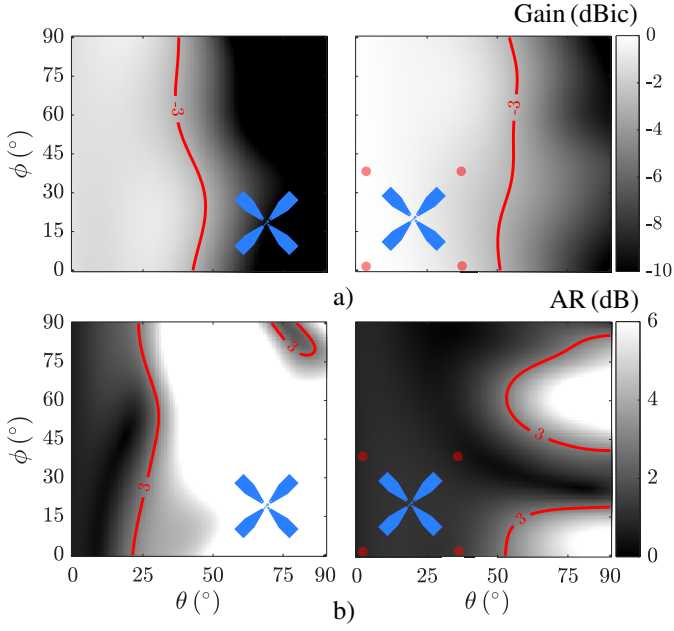

The pin contribution is interpreted as a steering-dependent inductive reactance
\begin{equation}
\label{eq:active_reactance}
X_\mathrm{A}(\theta_0)=\omega L(\theta_0)
\end{equation}
where the equivalent inductance is estimated by a modified Grover-type expression~\cite{Grover2004},
\begin{equation}
\label{eq:grover_inductance}
L_p(\theta_0)=\frac{\mu_0}{2\pi}l_{\mathrm{eff}}(\theta_0)
\left[
\ln\left(
\frac{2l_{\mathrm{eff}}(\theta_0)}
{w_{\mathrm{eff}}(\theta_0)}
\right)-C
\right].
\end{equation}

\noindent Here, \(C\) is an empirical correction factor accounting for the pin current distribution and finite conductor geometry. The effective dimensions are obtained from
\begin{align}
&l_\mathrm{eff}(\theta_0) = \max\{v_L(\theta_0),v_W(\theta_0)\},\\
&w_\mathrm{eff}(\theta_0) = \min\{v_L(\theta_0),v_W(\theta_0)\},\\
&v_L(\theta_0) = h_p P_L^p(\theta_0)+w_p\left[1-P_L^p(\theta_0)\right], \text{and}\\
&v_W(\theta_0) = w_p P_W^p(\theta_0)+h_p\left[1-P_W^p(\theta_0)\right],
\end{align}

\noindent with \(P_L(\theta_0)=|\cos\theta_0|\) and \(P_W(\theta_0)=|\sin\theta_0|\). These projection terms provide a compact way to describe how the tangential electric field encounters the pin geometry as the beam is steered. Near broadside, the field mainly interacts with the narrow transverse dimension of the pin, whereas E-plane steering increases the effective interaction with the vertical pin height. The dominance factor \(p\) controls how rapidly this transition occurs between the two limiting dimensions. Thus, the pins are modeled as reactive loading structures whose effective inductance increases with the scan angle, compensating the capacitive active-impedance shift of the unloaded dipole. The active impedance in Fig.~\ref{fig:XB_1x3_A}a confirms that the pins mainly compensate the scan-dependent reactance, while the compact-model comparison in Fig.~\ref{fig:XB_1x3_A}b captures this trend using \(C=0.75\) and \(p=7\) for \(h_p=3\,\text{mm}\) and \(h_p=6\,\text{mm}\), and \(C=0.6\) and \(p=3\) for \(h_p=9\,\text{mm}\). The resulting active reflection coefficient is improved over the steering range in Fig.~\ref{fig:XB_1x3_A}c and across frequency in Fig.~\ref{fig:XB_1x3_A}d. Since \(h_p=6\,\text{mm}\) provides the most consistent scan-dependent reactance compensation and the lowest active-reflection-coefficient variation over both steering angle and frequency, it is selected for the final X-band element.

Finally, the dual-polarized X-band unit cell is replicated into a \(5\times5\) finite array with the non-excited ports terminated in \(50~\Omega\). Compared with the original element, the pin-enhanced element provides a wider embedded CP gain response in Fig.~\ref{fig:XB_5x5}a and satisfies the baseline 3-dB AR beamwidth in Fig.~\ref{fig:XB_5x5}b, which is required before the polarization-synthesis stage.

\subsection{S-Band Element}\label{sec:SB}

\begin{figure}[t]
\centering
\def\svgwidth{240pt}
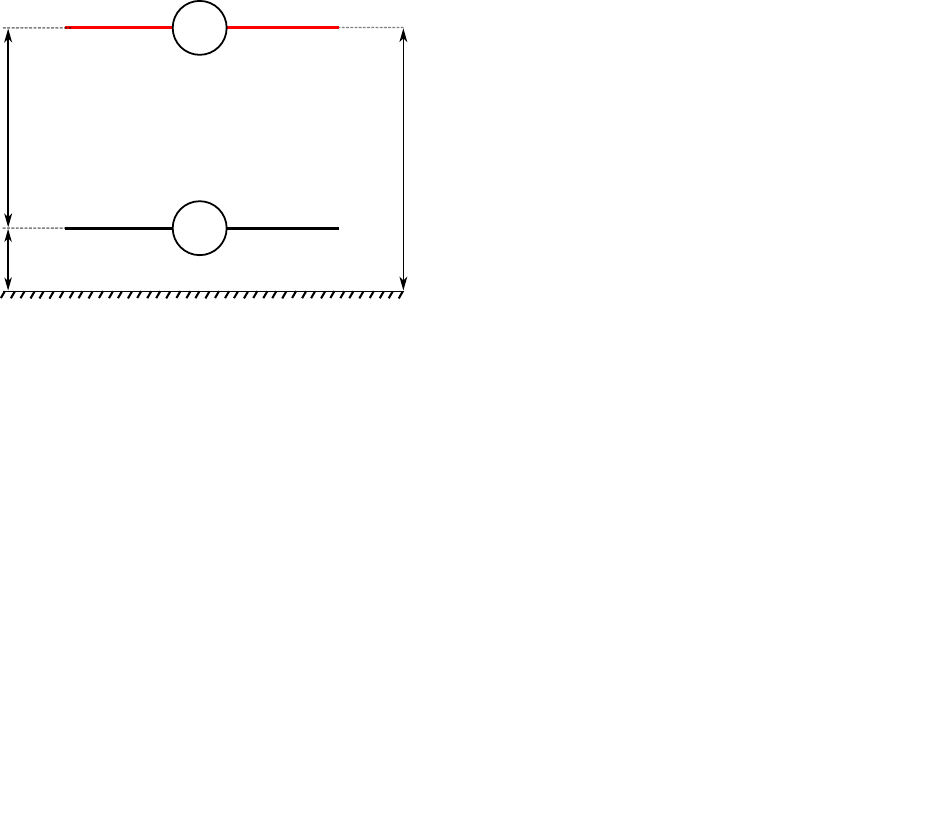
\caption{Theoretical and simulated inter-band coupling analysis: a) simplified two-dipole setup, b) coupling as a function of normalized vertical separation \(h_{\mathrm{S,X}}/\lambda_\mathrm{X}\), and c) estimated theoretical coupling as a function of the S-band radiator reflection coefficient at X-band for different separations.}
\label{fig:C_SX_theory}
\end{figure}

The S-band radiator must provide the S-band embedded element response while suppressing coupling to the X-band subarray below it. During early iterations, this tradeoff is studied with the theoretical coupling model in Fig.~\ref{fig:C_SX_theory}a rather than by repeatedly meshing the full S-/X-band geometry.

The S--X interaction is approximated by a direct term and a reflector-image term,
\begin{equation}
    Z_{\mathrm{S,X}}^\mathrm{tot}
    =
    Z_{\mathrm{S,X}}^\mathrm{dir}(r_\mathrm{dir})
    -
    Z_{\mathrm{S,X}}^\mathrm{img}(r_\mathrm{img})
\end{equation}
where \(r_\mathrm{dir}=h_{\mathrm{S,X}}\) and \(r_\mathrm{img}=2h_\mathrm{X}+h_{\mathrm{S,X}}\). Each contribution is estimated using the three-term Hertzian-dipole field expansion~\cite{Balanis2016}:
\begin{equation}
\label{eq:ZSX_three_term}
    Z_{\mathrm{S,X}}(r)
    \approx
    60e^{-jkr}\left[
    \frac{1}{kr}
    +j\frac{1}{(kr)^2}
    -j\frac{1}{(kr)^3}
    \right]
    \qquad (\Omega)
\end{equation}
where \(k=2\pi/\lambda_\mathrm{X}\). The approximate coupling coefficient is
\begin{equation}
\label{eq:CSX_theor}
    C_{\mathrm{S,X}}
    \approx
    \frac{Z_{\mathrm{S,X}}^\mathrm{tot}}{2Z_0}
    \sqrt{1-\left|\Gamma_\mathrm{S}(f_\mathrm{X})\right|^2}.
\end{equation}

\begin{figure}[t]
\centering
\def\svgwidth{240pt}
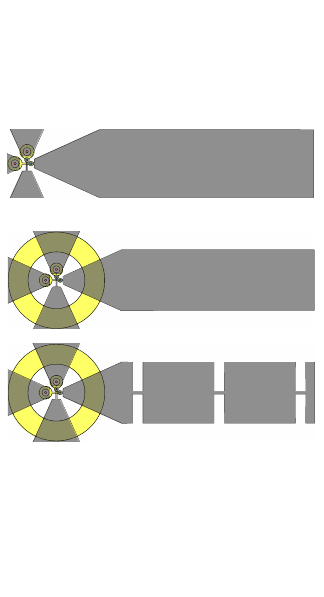
\caption{Evolution of the S-band radiator: a) initial dipole and corresponding worst-case S-band active reflection coefficient and theoretical X-band coupling, b) microstrip-ring-loaded dipole and the aforementioned parameters as a function of ring radius and dipole extension, and c) final slotted dipole with microstrip ring maximum active reflection coefficient and maximum inter-band coupling as a function of $W_\text{i}$ and $W_\text{g}$.}
\label{fig:SB_evo}
\end{figure}

\begin{figure}[t]
\centering
\def\svgwidth{250pt}
%% Creator: Inkscape 1.4.3 (0d15f75, 2025-12-25), www.inkscape.org
%% Accompanies image file 'S_AEPs_freqs.pdf' (pdf, eps, ps)
\begingroup%
  \makeatletter%
  \providecommand\color[2][]{%
    \errmessage{(Inkscape) Color is used for the text in Inkscape, but the package 'color.sty' is not loaded}%
    \renewcommand\color[2][]{}%
  }%
  \providecommand\transparent[1]{%
    \errmessage{(Inkscape) Transparency is used (non-zero) for the text in Inkscape, but the package 'transparent.sty' is not loaded}%
    \renewcommand\transparent[1]{}%
  }%
  \providecommand\rotatebox[2]{#2}%
  \newcommand*\fsize{\dimexpr\f@size pt\relax}%
  \newcommand*\lineheight[1]{\fontsize{\fsize}{#1\fsize}\selectfont}%
  \ifx\svgwidth\undefined%
    \setlength{\unitlength}{426.55690591bp}%
    \ifx\svgscale\undefined%
      \relax%
    \else%
      \setlength{\unitlength}{\unitlength * \real{\svgscale}}%
    \fi%
  \else%
    \setlength{\unitlength}{\svgwidth}%
  \fi%
  \global\let\svgwidth\undefined%
  \global\let\svgscale\undefined%
  \makeatother%
  \begin{picture}(1,0.951432)%
    \lineheight{1}%
    \setlength\tabcolsep{0pt}%
    \put(0,0){\includegraphics[width=\unitlength,page=1]{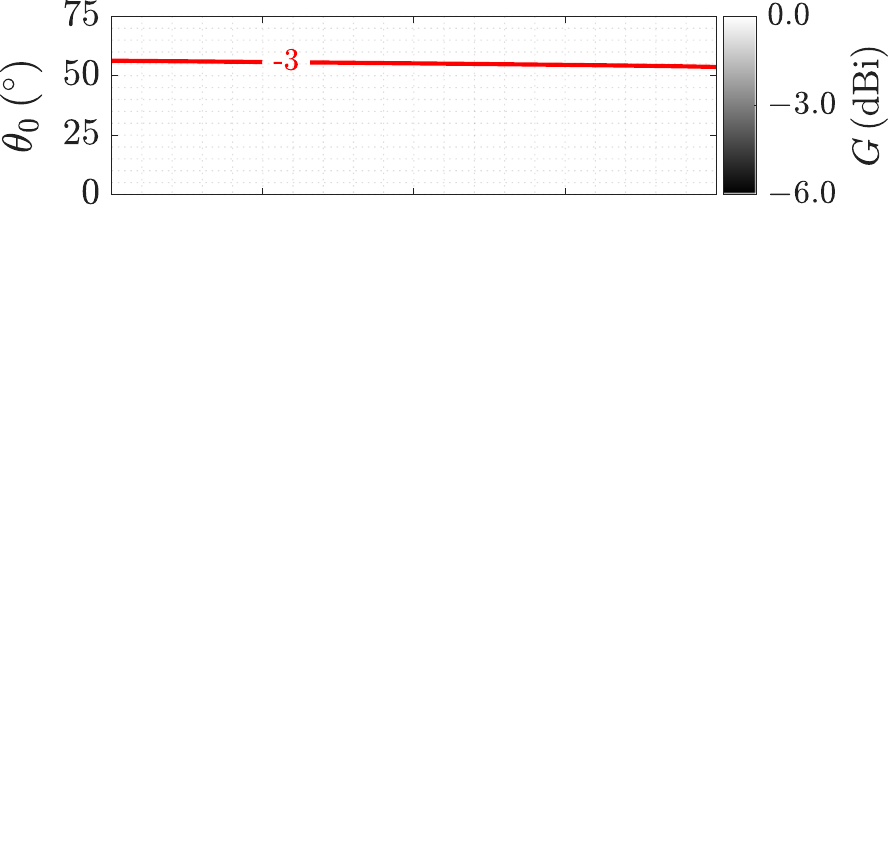}}%
    \put(0.47228806,0.68004811){\color[rgb]{0,0,0}\makebox(0,0)[t]{\lineheight{1.25}\smash{\begin{tabular}[t]{c}a)\end{tabular}}}}%
    \put(0.47198912,0.40036583){\color[rgb]{0,0,0}\makebox(0,0)[t]{\lineheight{1.25}\smash{\begin{tabular}[t]{c}b)\end{tabular}}}}%
    \put(0.47189707,0.00715465){\color[rgb]{0,0,0}\makebox(0,0)[t]{\lineheight{1.25}\smash{\begin{tabular}[t]{c}c)\end{tabular}}}}%
    \put(0,0){\includegraphics[width=\unitlength,page=2]{S_AEPs_freqs.pdf}}%
  \end{picture}%
\endgroup%

\caption{S-band dipole normalized active element patterns across frequency for a) \(\phi_0=0^\circ\), b) \(\phi_0=45^\circ\), and c) \(\phi_0=90^\circ\).}
\label{fig:S_AEP_Freq}
\end{figure}

Thus, the S-band radiator must radiate efficiently in the S-band while presenting a high-reflection, filtering response at X-band. The model agrees well with simulation for the intended separation in Fig.~\ref{fig:C_SX_theory}b, and shows that X-band mismatch of the S-band radiator is a dominant parameter for isolating collinear dipoles in Fig.~\ref{fig:C_SX_theory}c. The analysis is carried out at X-band frequencies, where higher-order resonant behavior of the larger S-band dipole is more 
likely to overlap with the X-band operation than in the opposite case.

The S-band design evolution is summarized in Fig.~\ref{fig:SB_evo}. The initial dipole in Fig.~\ref{fig:SB_evo}a cannot satisfy the inter-band isolation requirement. Therefore, it is first loaded with a microstrip ring as shown in Fig.~\ref{fig:SB_evo}b, which adds S-band inductance and increases the S-band radiator reflection at X-band, yielding an extended solution space. Slots are then introduced into the dipole arms to strengthen the low-pass filtering response, as shown in Fig.\ref{fig:SB_evo}c. The selected dimensions, \(E_\mathrm{S}=19\,\text{mm}\), \(r_\mathrm{o}=4.75\,\text{mm}\), \(W_\mathrm{i}=0.3\,\text{mm}\), and \(W_\mathrm{g}=1~\mathrm{mm}\), balance active matching, isolation margin, and manufacturability. A related inter-band filtering dipole was previously used as a single element in a simulation study~\cite{SAA_FSS_1}. In the present work, this filtering principle is experimentally validated in array prototypes.

The final S-band active element patterns are shown in Fig.~\ref{fig:S_AEP_Freq}. The normalized CP gain maintains a \(-3\)-dB beamwidth exceeding the target \(\pm50^\circ\) scan range over the receiving frequencies, providing the required baseline response before calibrated polarization synthesis.

\subsection{S-/X-Band Unit-Cell Characterization}

\begin{figure}[t]
\centering
\def\svgwidth{250pt}
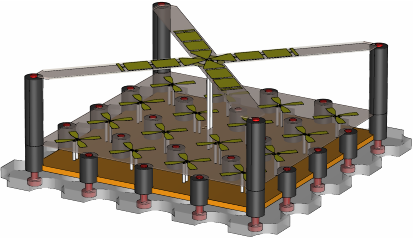
\caption{Proposed dual-band SAPAA unit cell.}
\label{fig:SX_UC}
\end{figure}

The final shared-aperture unit cell is shown in Fig.~\ref{fig:SX_UC}. The X-band PCB, consisting of 16 sequentially rotated dual-polarized elements, is placed below the S-band radiator. Both the S-band and X-band PCBs are mounted above the copper reflector using dielectric supports and M2 screws. The screw height determines the effective X-band inductive loading, as was discussed in Section~II-A. Simulations with periodic boundary conditions are used to evaluate the large-array behavior.

\begin{figure}[t]
\centering
\def\svgwidth{250pt}
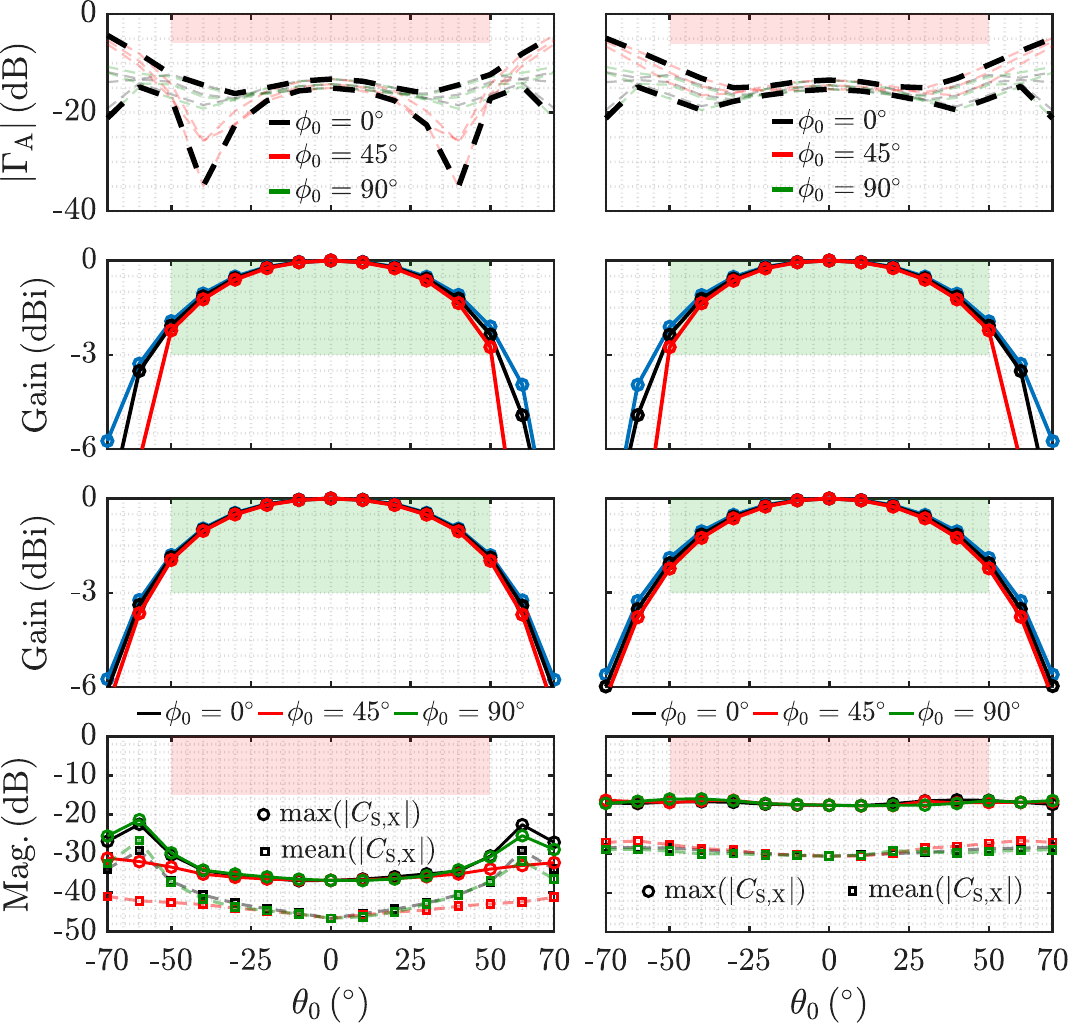
\caption{S-band element performance of the SAPAA unit cell for port 1 (left) and port 2 (right): a) active reflection coefficient, b) active element pattern in the \(\phi_0=0^\circ\) plane, c) active element pattern in the \(\phi_0=45^\circ\) plane, and d) maximum and average coupling between S- and X-band ports in the S-band (left) and X-band (right).}
\label{fig:SX_UC_S}
\end{figure}

The S-band results in Fig.~\ref{fig:SX_UC_S} show that the active reflection coefficients of both S-band ports remain below \(-10\)\,dB over the target band and scan range. The active element patterns satisfy the \(\pm50^\circ\) requirement in the principal plane, whereas the diagonal plane is wider because of the smaller projected spacing. The inter-band coupling remains low, the worst-case S-band coupling is \(-29.8\)\,dB and the worst-case X-band coupling is \(-16\)\,dB, satisfying the \(-15\)-dB target and agreeing well with the analytical approximation.

\begin{figure}[t]
\centering
\def\svgwidth{250pt}
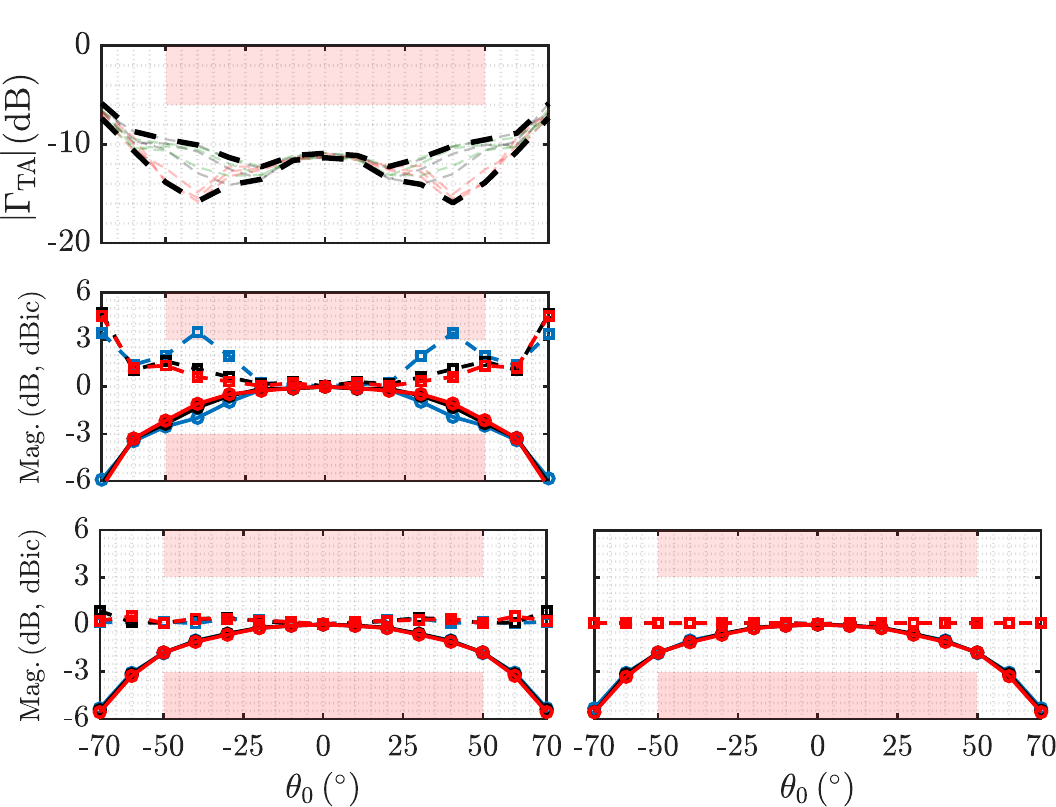
\caption{X-band array performance for fixed-polarization baseline excitations (left) and per-element AR-constrained excitations (right): a) total active reflection coefficient, b) scan gain and axial-ratio envelopes for \(\phi_0=0^\circ\), and c) scan gain and axial-ratio envelopes for \(\phi_0=45^\circ\).}
\label{fig:SX_UC_X}
\end{figure}

The X-band \(4\times4\) subarray is evaluated through its total active reflection coefficient, SGE, and ARE, as shown in Fig.~\ref{fig:SX_UC_X}. Two excitation strategies are compared, a fixed-polarization baseline and the proposed per-element AR-constrained excitation. For a given frequency \(f\) and steering direction \((\theta_0,\phi_0)\), the \(\theta\)- and \(\phi\)-polarized embedded field samples of the \(N\) active ports are collected into the vectors
\begin{equation}
\label{eq:embedded_field_vectors}
    \mathbf{E}_{\theta}(\theta_0,\phi_0,f)
    =
    \begin{bmatrix}
    E_{\theta,1} & E_{\theta,2} & \cdots & E_{\theta,N}
    \end{bmatrix}^{T},
\end{equation}
\begin{equation}
    \mathbf{E}_{\phi}(\theta_0,\phi_0,f)
    =
    \begin{bmatrix}
    E_{\phi,1} & E_{\phi,2} & \cdots & E_{\phi,N}
    \end{bmatrix}^{T}.
\end{equation}

\noindent \(\mathbf{E}_{\theta},\mathbf{E}_{\phi}\in\mathbb{C}^{N\times1}\) are obtained from the embedded-element field matrices by selecting the samples corresponding to the intended steering direction. In the unconstrained-AR baseline case, the receive excitation vector is obtained from
\begin{equation}
\label{eq:baseline_excitation}
    \mathbf{a}_0
    =
    \mathbf{E}_{\theta}^{*}p_{\theta,0}^{*}
    +
    \mathbf{E}_{\phi}^{*}p_{\phi,0}^{*},
    \qquad
    \mathbf{a}_0\in\mathbb{C}^{N\times1}.
\end{equation}
Here, the same target polarization vector is used for all active ports,
\begin{equation}
\label{eq:fixed_pol_vector}
    \mathbf{p}_0
    =
    \begin{bmatrix}
    p_{\theta,0} \\
    p_{\phi,0}
    \end{bmatrix}
    =
    \begin{bmatrix}
    1 \\
    e^{j\Delta\psi_0}
    \end{bmatrix},
\end{equation}
\noindent where \(p_{\theta,0}\) and \(p_{\phi,0}\) are scalar polarization weights and \(\Delta\psi_0=90^\circ\) for the baseline circular-polarization case, according to the adopted handedness convention. Thus, each active port obtains a different complex coefficient through its embedded field response, but the target polarization phase \(\Delta\psi_0\) is common for all active ports. This fixed-phase baseline provides acceptable scan gain and active matching. However, the axial-ratio envelope reaches 3.5\,dB at \(\theta_0 = 40^\circ\) in the principal plane, indicating that a common polarization phase does not fully compensate the embedded amplitude and phase imbalance of the integrated shared aperture.

To reduce the residual polarization imbalance, the excitation vector is selected by maximizing the gain in the desired circular-polarization state under an AR constraint. For each frequency and steering direction, the selected excitation vector is defined as
\begin{equation}
\label{eq:ar_threshold_selection}
    \mathbf{a}_\mathrm{opt}
    =
    \arg\max_{\mathbf{a}\in\mathcal{A}}
    \left\{
    G_{\chi}(\mathbf{a}) : AR(\mathbf{a}) \leq AR_\mathrm{max}
    \right\},
\end{equation}
\noindent with \(AR_\mathrm{max}=0.1\)\,dB. Here, \(\chi\in\{\mathrm{RHCP},\mathrm{LHCP}\}\) denotes the selected co-polar receive state, \(\mathcal{A}\) is the set of evaluated candidate excitation vectors, and \(G_{\chi}(\mathbf{a})\) and \(AR(\mathbf{a})\) are the gain and axial ratio obtained with the candidate excitation vector \(\mathbf{a}\). If no candidate satisfies the threshold, the minimum-AR candidate is selected, with \(G_{\chi}\) used as the secondary criterion. 

In the proposed per-element case, the candidate set \(\mathcal{A}\) includes vectors $\bm{a}_q \in \mathbb{C}^{N}$ with elements

\label{eq:per_element_coefficient}
\begin{equation}
    a_{n,q_n}
    =
    E_{\theta,n}^{*}p_{\theta,q_n}^{*}
    +
    E_{\phi,n}^{*}p_{\phi,q_n}^{*}.
\end{equation}

\noindent The element-dependent polarization vector is now 

\label{eq:per_element_pol_vector}
\begin{equation}
    \mathbf{p}_{q_n}
    =
    \begin{bmatrix}
    p_{\theta,q_n} \\
    p_{\phi,q_n}
    \end{bmatrix}
    =
    \begin{bmatrix}
    1 \\
    e^{j\Delta\psi_{q_n}}
    \end{bmatrix}
\end{equation}

\noindent where

\label{eq:per_element_phase_selection}
\begin{equation}
    \Delta\psi_{q_n}
    =
    \Delta\psi_{q_n}(\theta_0,\phi_0,f),
    \qquad n=1,\ldots,N.
\end{equation}

\noindent Thus, the primary optimization variables are the phases $\Delta\psi\in[0,2\pi]$. Consequently, the proposed per-element excitation is no longer constrained to use an identical relative polarization phase for all ports. Instead, the selected phase can vary with element, frequency, and steering direction, which compensates residual embedded amplitude and phase imbalance over frequency and scan angle. The right side of Fig.~\ref{fig:SX_UC_X} shows that this method reduces the worst-case AR below 0.1\,dB while preserving scan gain and acceptable total active reflection coefficient.

\begin{figure}[t]
\centering
\def\svgwidth{250pt}
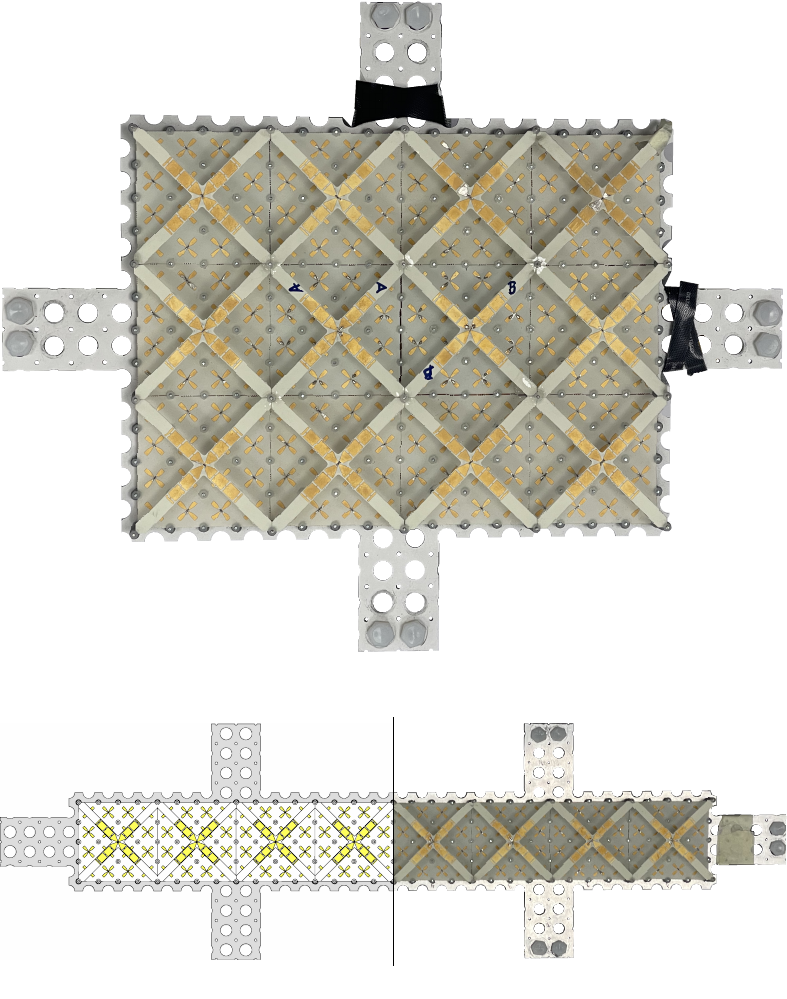
\caption{Fabricated prototypes: a) \(3\times4\) SAPAA, where the two center SAPAA unit cells are fully cabled and the surrounding unit cells are terminated with \(50~\Omega\) loads, and b) \(1\times8\) SAPAA, where the S-band elements are fully cabled and the X-band elements are terminated.}
\label{fig:protos}
\end{figure}

\begin{figure}[t]
\centering
\def\svgwidth{250pt}
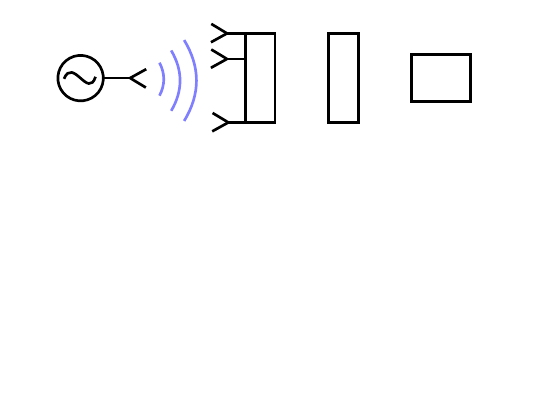
\caption{Active measurement setups for a) the \(1\times8\) prototype and b) the \(3\times4\) prototype.}
\label{fig:setups}
\end{figure}

\section{Experimental Validation}\label{ch:experimental}

Fig.~\ref{fig:protos} shows the two prototypes realized from the same SAPAA unit cell, and their active measurement setups are illustrated in Fig.~\ref{fig:setups}. The \(3\times4\) SAPAA prototype has two fully cabled central unit cells, resulting in 68 accessible ports, while the surrounding unit cells are terminated to emulate the loading environment of a larger array. It is used to validate passive matching, inter-band isolation, and X-band beamforming with the AR-constrained excitations. The \(1\times8\) prototype has fully cabled S-band elements and terminated X-band elements and is used to validate S-band beamforming and polarization synthesis. The fabricated holder was manufactured without the intended M2 thread. Therefore, nylon nuts were used, causing the screw ends to extend slightly above the antenna plane. This mainly affects the X-band performance at large scan angles, as was studied in \ref{sec:XB}.

All radiation measurements were performed in an anechoic chamber using two approaches. In the passive measurements, each embedded response was measured sequentially with the remaining ports terminated, and array beams were reconstructed from the measured fields and selected excitation coefficients. In the RFSoC-based active measurements, all available receive channels were measured simultaneously using an eight-channel Zynq UltraScale+ RFSoC beamforming platform. The active S-band measurements were performed by connecting the prototype directly to the RFSoC, whereas the X-band active measurements used external down-conversion because the RFSoC platform operates up to 6\,GHz.

\subsection{\(3\times4\) Array Prototype}

\begin{figure}[t]
\centering
\def\svgwidth{250pt}
%% Creator: Inkscape 1.4.3 (0d15f75, 2025-12-25), www.inkscape.org
%% Accompanies image file 'SX_3x4_passives_V3.pdf' (pdf, eps, ps)
\begingroup%
  \makeatletter%
  \providecommand\color[2][]{%
    \errmessage{(Inkscape) Color is used for the text in Inkscape, but the package 'color.sty' is not loaded}%
    \renewcommand\color[2][]{}%
  }%
  \providecommand\transparent[1]{%
    \errmessage{(Inkscape) Transparency is used (non-zero) for the text in Inkscape, but the package 'transparent.sty' is not loaded}%
    \renewcommand\transparent[1]{}%
  }%
  \providecommand\rotatebox[2]{#2}%
  \newcommand*\fsize{\dimexpr\f@size pt\relax}%
  \newcommand*\lineheight[1]{\fontsize{\fsize}{#1\fsize}\selectfont}%
  \ifx\svgwidth\undefined%
    \setlength{\unitlength}{595.00067283bp}%
    \ifx\svgscale\undefined%
      \relax%
    \else%
      \setlength{\unitlength}{\unitlength * \real{\svgscale}}%
    \fi%
  \else%
    \setlength{\unitlength}{\svgwidth}%
  \fi%
  \global\let\svgwidth\undefined%
  \global\let\svgscale\undefined%
  \makeatother%
  \begin{picture}(1,0.65703053)%
    \lineheight{1}%
    \setlength\tabcolsep{0pt}%
    \put(0.51613224,0.35583396){\color[rgb]{0,0,0}\makebox(0,0)[t]{\lineheight{1.25}\smash{\begin{tabular}[t]{c}a)\end{tabular}}}}%
    \put(0.51614326,0.00548644){\color[rgb]{0,0,0}\makebox(0,0)[t]{\lineheight{1.25}\smash{\begin{tabular}[t]{c}b)\end{tabular}}}}%
    \put(0,0){\includegraphics[width=\unitlength,page=1]{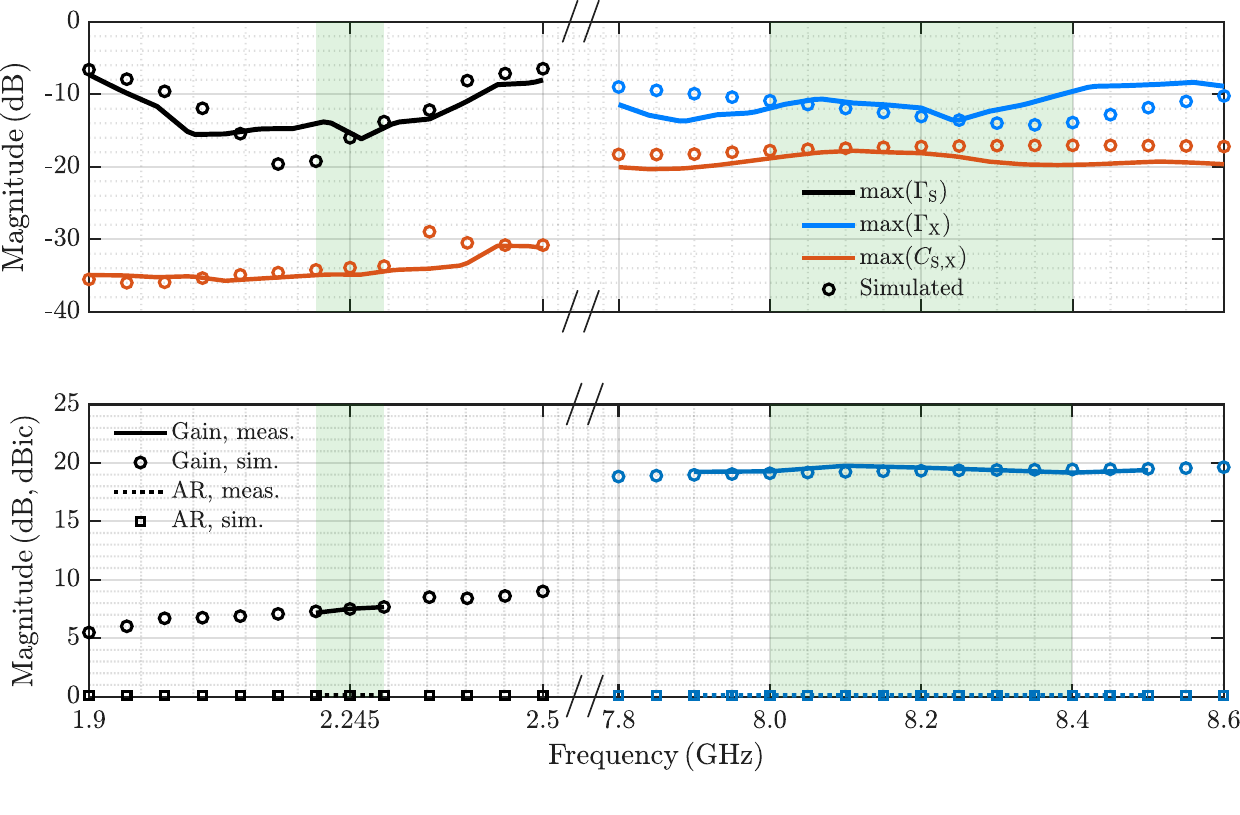}}%
  \end{picture}%
\endgroup%

\caption{\(3\times4\) prototype S-band performance, where the S-band utilizes the \(1\times2\) array configuration and the X-band the \(4\times8\) configuration, respectively. a) simulated and measured worst-case passive reflection coefficients and inter-band isolation, and b) simulated and measured broadside gains and axial ratios for S-band and X-band operation.}
\label{fig:3x4_passives}
\end{figure}

The passive reflection and inter-band-coupling envelopes of the \(3\times4\) prototype are shown in Fig.~\ref{fig:3x4_passives}a. The measured results agree well with the simulation. Both bands satisfy the 10-dB passive matching criterion, and the measured inter-band isolation is better than 17\,dB. Fig.~\ref{fig:3x4_passives}b shows the broadside gain and AR obtained using the AR-constrained excitation coefficients. The simulated and measured broadside gains agree closely, with average differences of 0.04\,dB in the S-band and 0.13\,dB in the X-band, while the AR remains below 0.1\,dB in both bands.

\begin{figure}[t]
\centering
\def\svgwidth{250pt}
%% Creator: Inkscape 1.4.3 (0d15f75, 2025-12-25), www.inkscape.org
%% Accompanies image file 'S_3x4_SGE_ARE_TARC_normalized.pdf' (pdf, eps, ps)
\begingroup%
  \makeatletter%
  \providecommand\color[2][]{%
    \errmessage{(Inkscape) Color is used for the text in Inkscape, but the package 'color.sty' is not loaded}%
    \renewcommand\color[2][]{}%
  }%
  \providecommand\transparent[1]{%
    \errmessage{(Inkscape) Transparency is used (non-zero) for the text in Inkscape, but the package 'transparent.sty' is not loaded}%
    \renewcommand\transparent[1]{}%
  }%
  \providecommand\rotatebox[2]{#2}%
  \newcommand*\fsize{\dimexpr\f@size pt\relax}%
  \newcommand*\lineheight[1]{\fontsize{\fsize}{#1\fsize}\selectfont}%
  \ifx\svgwidth\undefined%
    \setlength{\unitlength}{542.60146734bp}%
    \ifx\svgscale\undefined%
      \relax%
    \else%
      \setlength{\unitlength}{\unitlength * \real{\svgscale}}%
    \fi%
  \else%
    \setlength{\unitlength}{\svgwidth}%
  \fi%
  \global\let\svgwidth\undefined%
  \global\let\svgscale\undefined%
  \makeatother%
  \begin{picture}(1,0.32661141)%
    \lineheight{1}%
    \setlength\tabcolsep{0pt}%
    \put(0,0){\includegraphics[width=\unitlength,page=1]{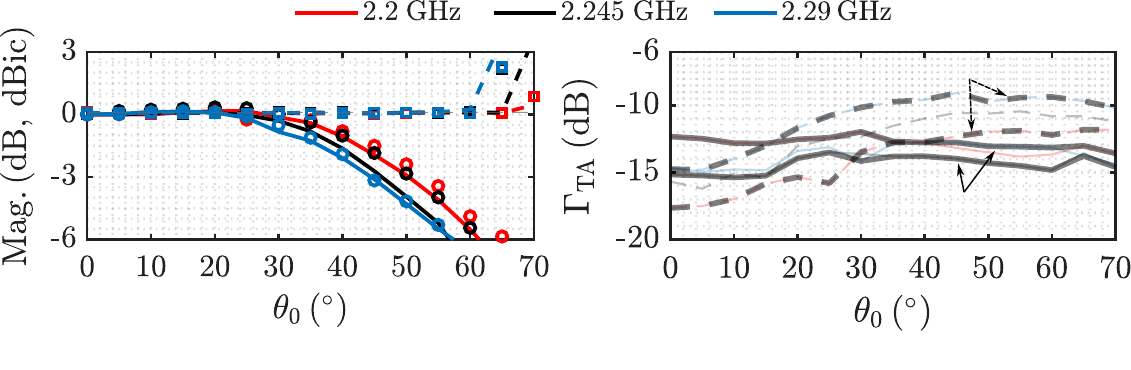}}%
    \put(0.86416962,0.14215183){\color[rgb]{0,0,0}\makebox(0,0)[t]{\lineheight{1.25}\smash{\begin{tabular}[t]{c}\tiny{meas. envelope}\end{tabular}}}}%
    \put(0.84840795,0.26145509){\color[rgb]{0,0,0}\makebox(0,0)[t]{\lineheight{1.25}\smash{\begin{tabular}[t]{c}\tiny{sim. envelope}\end{tabular}}}}%
    \put(0,0){\includegraphics[width=\unitlength,page=2]{S_3x4_SGE_ARE_TARC_normalized.pdf}}%
    \put(0.25196323,0.00455085){\color[rgb]{0,0,0}\makebox(0,0)[t]{\lineheight{1.25}\smash{\begin{tabular}[t]{c}a)\end{tabular}}}}%
    \put(0.76730059,0.00161362){\color[rgb]{0,0,0}\makebox(0,0)[t]{\lineheight{1.25}\smash{\begin{tabular}[t]{c}b)\end{tabular}}}}%
    \put(0,0){\includegraphics[width=\unitlength,page=3]{S_3x4_SGE_ARE_TARC_normalized.pdf}}%
  \end{picture}%
\endgroup%

\caption{\(1\times2\) S-band array performance of the \(3\times4\) array prototype. a) simulated and measured scan gain and axial-ratio envelopes with the 0.1-dB-AR-constrained excitations in the \(\phi_0=0^\circ\) scan plane, and b) simulated and measured total active reflection coefficient.}
\label{fig:3x4_actives_SB}
\end{figure}

The S-band scan results of the \(1\times2\) configuration inside the \(3\times4\) prototype are shown in Fig.~\ref{fig:3x4_actives_SB}a. The AR remains below 0.1\,dB over the evaluated scan range, the normalized SGEs agree within measurement accuracy, and the measured total active reflection coefficient remains below \(-12.5\)\,dB as shown in Fig.~\ref{fig:3x4_actives_SB}b. In recent measured phased-array prototypes, an active reflection coefficient below \(-6\)\,dB is commonly used as a practical reference for acceptable scan-active matching~\cite{Ullah2025WidebandConnectedSlotArrayFR1,Fan2025EdgeTruncationUWBPhasedArrays,Sun2023AllMetalFullPolarization,Kuosmanen2023FilteringCorrugatedVivaldi,Mu2024UWB_CP_PhasedArrayLowVSWR,Chang2024UWBExponentialCurveDipoles,Zhao2024UWBDualPolarizedTCDA}.

\begin{figure}[t]
\centering
\def\svgwidth{250pt}
%% Creator: Inkscape 1.4.3 (0d15f75, 2025-12-25), www.inkscape.org
%% Accompanies image file 'SX_3x4_SGE_ARE_TARC_normalized_V2.pdf' (pdf, eps, ps)
\begingroup%
  \makeatletter%
  \providecommand\color[2][]{%
    \errmessage{(Inkscape) Color is used for the text in Inkscape, but the package 'color.sty' is not loaded}%
    \renewcommand\color[2][]{}%
  }%
  \providecommand\transparent[1]{%
    \errmessage{(Inkscape) Transparency is used (non-zero) for the text in Inkscape, but the package 'transparent.sty' is not loaded}%
    \renewcommand\transparent[1]{}%
  }%
  \providecommand\rotatebox[2]{#2}%
  \newcommand*\fsize{\dimexpr\f@size pt\relax}%
  \newcommand*\lineheight[1]{\fontsize{\fsize}{#1\fsize}\selectfont}%
  \ifx\svgwidth\undefined%
    \setlength{\unitlength}{544.24232303bp}%
    \ifx\svgscale\undefined%
      \relax%
    \else%
      \setlength{\unitlength}{\unitlength * \real{\svgscale}}%
    \fi%
  \else%
    \setlength{\unitlength}{\svgwidth}%
  \fi%
  \global\let\svgwidth\undefined%
  \global\let\svgscale\undefined%
  \makeatother%
  \begin{picture}(1,0.54662521)%
    \lineheight{1}%
    \setlength\tabcolsep{0pt}%
    \put(0,0){\includegraphics[width=\unitlength,page=1]{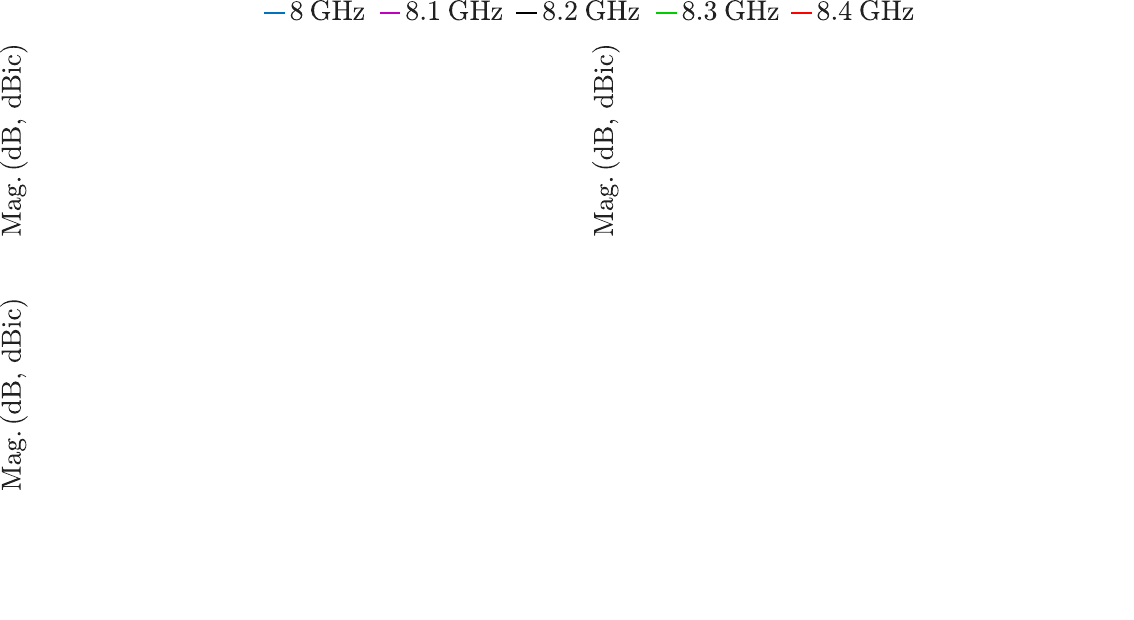}}%
    \put(0.26314878,0.00160877){\color[rgb]{0,0,0}\makebox(0,0)[t]{\lineheight{1.25}\smash{\begin{tabular}[t]{c}c)\end{tabular}}}}%
    \put(0.26318758,0.29607517){\color[rgb]{0,0,0}\makebox(0,0)[t]{\lineheight{1.25}\smash{\begin{tabular}[t]{c}a)\end{tabular}}}}%
    \put(0.78614934,0.29591331){\color[rgb]{0,0,0}\makebox(0,0)[t]{\lineheight{1.25}\smash{\begin{tabular}[t]{c}b)\end{tabular}}}}%
    \put(0.78614934,0.00160877){\color[rgb]{0,0,0}\makebox(0,0)[t]{\lineheight{1.25}\smash{\begin{tabular}[t]{c}d)\end{tabular}}}}%
    \put(0,0){\includegraphics[width=\unitlength,page=2]{SX_3x4_SGE_ARE_TARC_normalized_V2.pdf}}%
    \put(0.80268773,0.1233762){\color[rgb]{0,0,0}\makebox(0,0)[t]{\lineheight{1.25}\smash{\begin{tabular}[t]{c}min\end{tabular}}}}%
    \put(0.75197103,0.2402985){\color[rgb]{0,0,0}\makebox(0,0)[t]{\lineheight{1.25}\smash{\begin{tabular}[t]{c}max\end{tabular}}}}%
    \put(0,0){\includegraphics[width=\unitlength,page=3]{SX_3x4_SGE_ARE_TARC_normalized_V2.pdf}}%
  \end{picture}%
\endgroup%

\caption{\(4\times8\) X-band array performance of the \(3\times4\) prototype. Simulated and measured scan gain and axial-ratio envelopes with the AR-constrained excitations in the scan planes a) \(\phi_0=0^\circ\), b) \(\phi_0=45^\circ\), and c) \(\phi_0=90^\circ\) for the target X-band frequencies of 8, 8.1, 8.2, 8.3, and 8.4\,GHz. d) Simulated total active reflection coefficient for the evaluated cuts and frequencies, with the maximum and minimum envelopes.}
\label{fig:3x4_actives_XB_env}
\end{figure}

The X-band scan envelopes in Fig.~\ref{fig:3x4_actives_XB_env} show below-0.1-dB AR over the target band and scan range. The principal-plane scan gain agrees closely with simulation, whereas the diagonal-plane loss is slightly higher, consistent with the modified screw mounting and support-material's permittivity uncertainty. The simulated total active reflection coefficient remains below the \(-6\)-dB reference level. 

\begin{figure}[t]
\centering
\def\svgwidth{250pt}
%% Creator: Inkscape 1.4.3 (0d15f75, 2025-12-25), www.inkscape.org
%% Accompanies image file 'SX_3x4_X_patterns_Optim_V3.pdf' (pdf, eps, ps)
\begingroup%
  \makeatletter%
  \providecommand\color[2][]{%
    \errmessage{(Inkscape) Color is used for the text in Inkscape, but the package 'color.sty' is not loaded}%
    \renewcommand\color[2][]{}%
  }%
  \providecommand\transparent[1]{%
    \errmessage{(Inkscape) Transparency is used (non-zero) for the text in Inkscape, but the package 'transparent.sty' is not loaded}%
    \renewcommand\transparent[1]{}%
  }%
  \providecommand\rotatebox[2]{#2}%
  \newcommand*\fsize{\dimexpr\f@size pt\relax}%
  \newcommand*\lineheight[1]{\fontsize{\fsize}{#1\fsize}\selectfont}%
  \ifx\svgwidth\undefined%
    \setlength{\unitlength}{571.21847318bp}%
    \ifx\svgscale\undefined%
      \relax%
    \else%
      \setlength{\unitlength}{\unitlength * \real{\svgscale}}%
    \fi%
  \else%
    \setlength{\unitlength}{\svgwidth}%
  \fi%
  \global\let\svgwidth\undefined%
  \global\let\svgscale\undefined%
  \makeatother%
  \begin{picture}(1,0.69318006)%
    \lineheight{1}%
    \setlength\tabcolsep{0pt}%
    \put(0,0){\includegraphics[width=\unitlength,page=1]{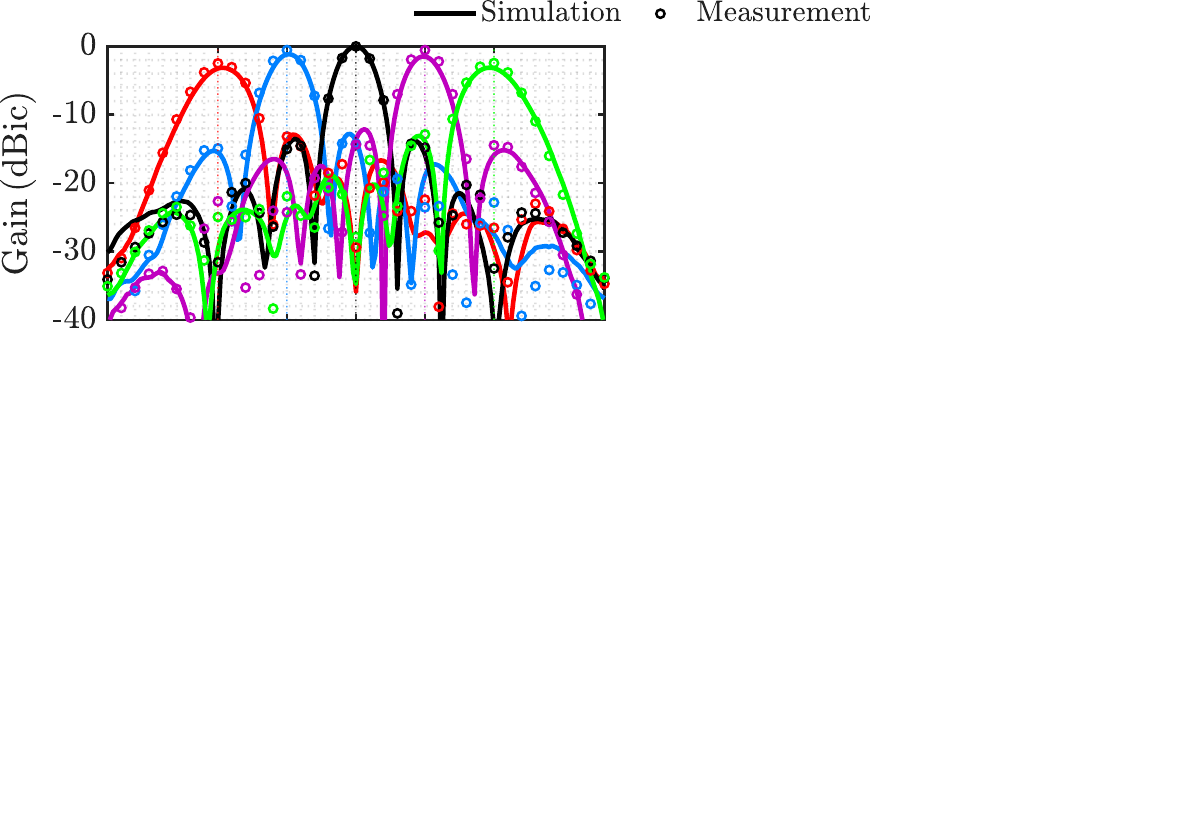}}%
    \put(0.29127147,0.00228457){\color[rgb]{0,0,0}\makebox(0,0)[t]{\lineheight{1.25}\smash{\begin{tabular}[t]{c}c)\end{tabular}}}}%
    \put(0.7687551,0.00228457){\color[rgb]{0,0,0}\makebox(0,0)[t]{\lineheight{1.25}\smash{\begin{tabular}[t]{c}d)\end{tabular}}}}%
    \put(0.29127147,0.37339224){\color[rgb]{0,0,0}\makebox(0,0)[t]{\lineheight{1.25}\smash{\begin{tabular}[t]{c}a)\end{tabular}}}}%
    \put(0.7687551,0.37339224){\color[rgb]{0,0,0}\makebox(0,0)[t]{\lineheight{1.25}\smash{\begin{tabular}[t]{c}b)\end{tabular}}}}%
    \put(0,0){\includegraphics[width=\unitlength,page=2]{SX_3x4_X_patterns_Optim_V3.pdf}}%
  \end{picture}%
\endgroup%

\caption{Measured and simulated radiation patterns of the \(4\times8\) X-band array array of the \(3\times4\) prototype using 0.1-dB-AR-constrained excitation coefficients for the cuts a) \(\phi=0^\circ\), b) \(\phi=45^\circ\), c) \(\phi=90^\circ\), and d) \(\phi=135^\circ\).}
\label{fig:3x4_X_patterns}
\end{figure}

\begin{figure}[t!]
\centering
\def\svgwidth{250pt}
%% Creator: Inkscape 1.4.3 (0d15f75, 2025-12-25), www.inkscape.org
%% Accompanies image file 'SX_3x4_X_2x2_RFSoC.pdf' (pdf, eps, ps)
\begingroup%
  \makeatletter%
  \providecommand\color[2][]{%
    \errmessage{(Inkscape) Color is used for the text in Inkscape, but the package 'color.sty' is not loaded}%
    \renewcommand\color[2][]{}%
  }%
  \providecommand\transparent[1]{%
    \errmessage{(Inkscape) Transparency is used (non-zero) for the text in Inkscape, but the package 'transparent.sty' is not loaded}%
    \renewcommand\transparent[1]{}%
  }%
  \providecommand\rotatebox[2]{#2}%
  \newcommand*\fsize{\dimexpr\f@size pt\relax}%
  \newcommand*\lineheight[1]{\fontsize{\fsize}{#1\fsize}\selectfont}%
  \ifx\svgwidth\undefined%
    \setlength{\unitlength}{540.39529155bp}%
    \ifx\svgscale\undefined%
      \relax%
    \else%
      \setlength{\unitlength}{\unitlength * \real{\svgscale}}%
    \fi%
  \else%
    \setlength{\unitlength}{\svgwidth}%
  \fi%
  \global\let\svgwidth\undefined%
  \global\let\svgscale\undefined%
  \makeatother%
  \begin{picture}(1,0.38060091)%
    \lineheight{1}%
    \setlength\tabcolsep{0pt}%
    \put(0,0){\includegraphics[width=\unitlength,page=1]{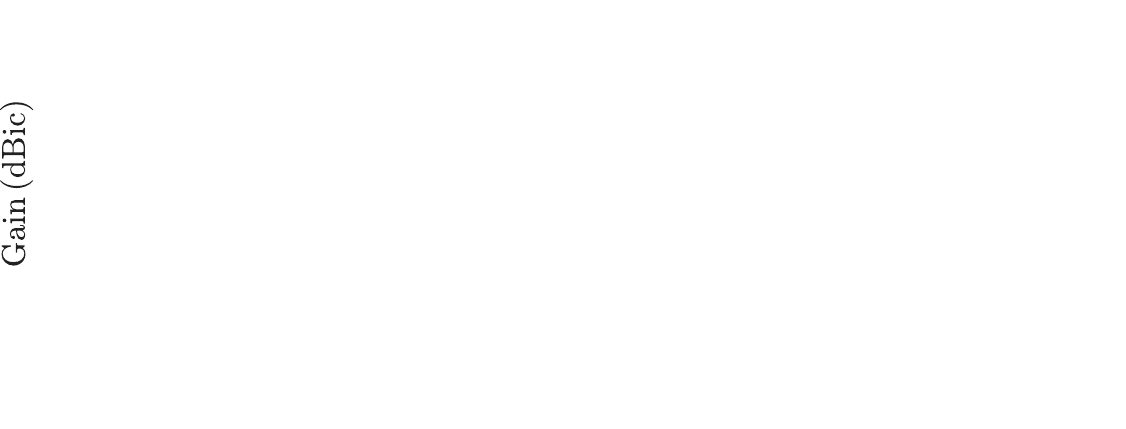}}%
    \put(0.29248719,0.00241492){\color[rgb]{0,0,0}\makebox(0,0)[t]{\lineheight{1.25}\smash{\begin{tabular}[t]{c}a)\end{tabular}}}}%
    \put(0.76918913,0.0024889){\color[rgb]{0,0,0}\makebox(0,0)[t]{\lineheight{1.25}\smash{\begin{tabular}[t]{c}b)\end{tabular}}}}%
    \put(0,0){\includegraphics[width=\unitlength,page=2]{SX_3x4_X_2x2_RFSoC.pdf}}%
  \end{picture}%
\endgroup%

\caption{Comparison of \(2\times2\) X-band center elements simulated, passive chamber measured, and active RFSoC measured broadside a) circularly polarized gain patterns and b) axial-ratio patterns at 8.2\,GHz for the \(\phi=0^\circ\) cut.}
\label{fig:3x4_2x2_X_patterns_RFSoC}
\end{figure}

Measured and simulated X-band radiation patterns with the AR-constrained coefficients are shown in Fig.~\ref{fig:3x4_X_patterns}. The main-beam directions and beam shapes agree well over the evaluated cuts. The passive and RFSoC-based X-band results are compared in Fig.~\ref{fig:3x4_2x2_X_patterns_RFSoC} for the central \(2\times2\) X-band elements, which match the eight available RFSoC channels. At 8.2\,GHz, the active pattern agrees moderately with the passive measurement and simulation, and the broadside AR remains below 0.1\,dB. The frequency-dependent RFSoC agreement was evaluated but is not shown for brevity. It was less consistent at 8\,GHz, but excellent agreement was observed at 8.4\,GHz. This behavior is likely caused by noise and conversion imbalance introduced by the external mixer chain. Direct X-band RFSoC operation or improved mixer-chain calibration is expected to improve this active-measurement agreement.

\subsection{\(1\times8\) Array Prototype}

\begin{figure}[t!]
\centering
\def\svgwidth{250pt}
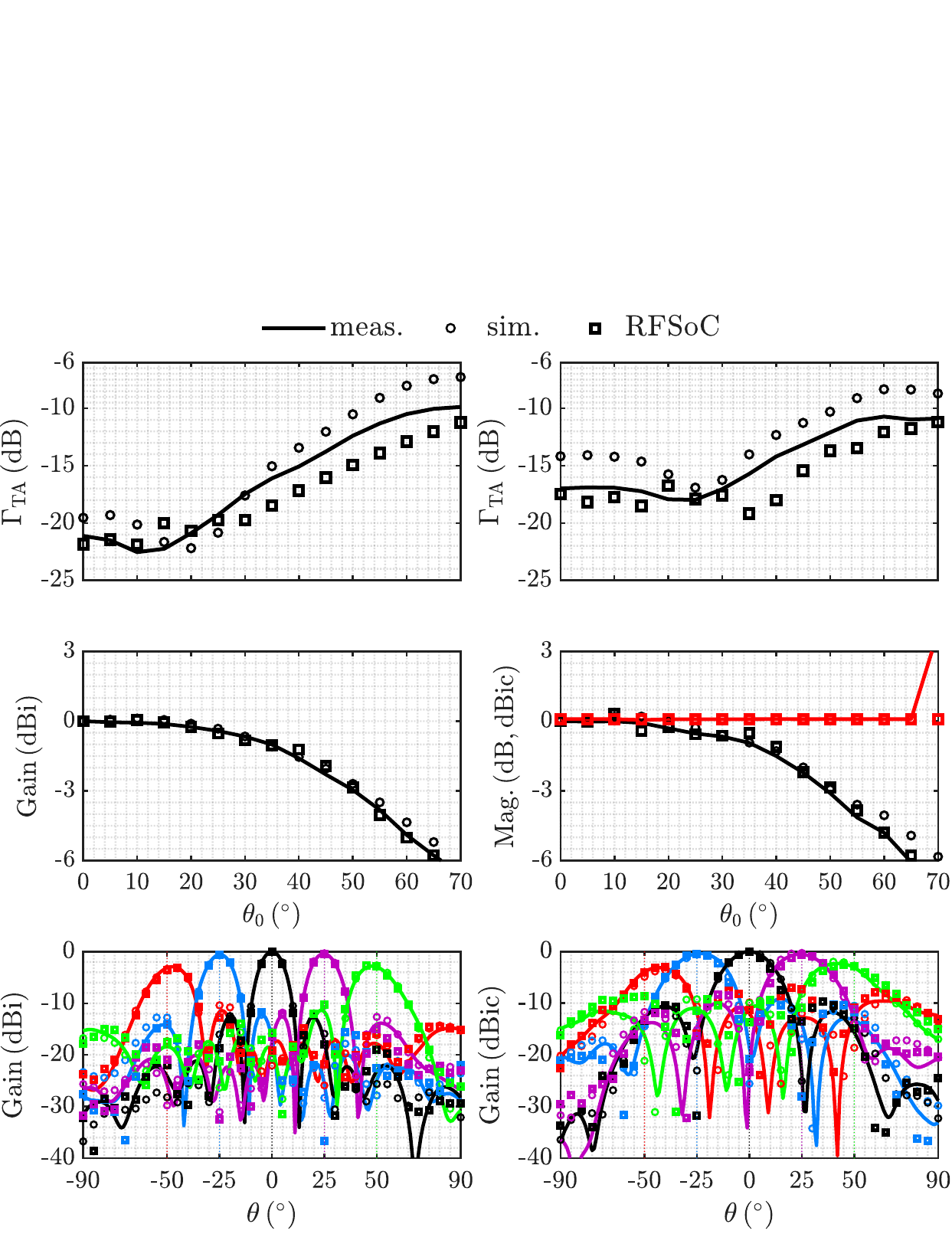
\caption{\(1\times8\) array prototype performance. The left column shows the linearly polarized results for the eight \(45^\circ\)-oriented ports, and the right column shows the circularly polarized results for the eight center dual-polarized ports: a) broadside realized gain across frequency, including axial ratio for the circularly polarized case, b) total active reflection coefficient at 2.245\,GHz, c) normalized scan gain envelope, including axial-ratio envelope for the circularly polarized case, and d) normalized radiation patterns at 2.245\,GHz.}
\label{fig:S_1x8_perf}
\end{figure}

The measured and simulated performance of the \(1\times8\) prototype is summarized in Fig.~\ref{fig:S_1x8_perf}. The left column presents the results for the eight linearly polarized \(45^\circ\)-oriented ports, whereas the right column presents the circularly polarized results obtained using the eight center dual-polarized ports.

Fig.~\ref{fig:S_1x8_perf}a shows the broadside realized gain as a function of frequency. The measured values are obtained from anechoic-chamber measurements and are compared with the corresponding full-wave simulations. For the linearly polarized case, the measured and simulated gains agree within 0.16\,dB over the evaluated frequency range, and the maximum measured gain is 14.38\,dBi. For the circularly polarized case, the maximum gain difference is 0.23\,dB, and the maximum measured gain is 11.32\,dBic. The circularly polarized axial ratio also remains below the specified target over the operating band.

The total active reflection coefficient at the center frequency of 2.245\,GHz is shown in Fig.~\ref{fig:S_1x8_perf}b as a function of steering angle. For both the linearly and circularly polarized cases, the measured total active reflection coefficient remains below \(-10\)\,dB across the evaluated steering range, which is well below the commonly used \(-6\)-dB limit for active phased arrays. Good agreement is observed between the simulation, passive anechoic-chamber measurement, and active RFSoC measurement.

The normalized SGE is shown in Fig.~\ref{fig:S_1x8_perf}c for both operating cases, together with the ARE for the circularly polarized case. The simulated, chamber-measured, and RFSoC-measured results follow the same scan-dependent trend within the target steering range. The circularly polarized ARE remains below 0.1\,dB, confirming that the selected excitation coefficients provide the intended polarization synthesis in the fabricated array.

The corresponding normalized radiation patterns at 2.245\,GHz are shown in Fig.~\ref{fig:S_1x8_perf}d. The measured main-beam directions and beam shapes agree well with the simulations for both measurement approaches. As expected, the linearly polarized \(1\times8\) configuration produces a narrower beam than the circularly polarized four center dual-polarized element configuration. These results validate the S-band beamforming performance of the proposed SAPAA prototype and confirm that the RFSoC-based active measurement captures the same array behavior observed in the passive measurements.

\section{Discussion}\label{ch:discussion}

\begin{table*}[t]
\centering
\caption{Comparison with state-of-the-art circularly polarized phased-array prototypes}
\label{tab:AR_comparison}
\renewcommand{\arraystretch}{1.08}

\resizebox{\textwidth}{!}{
\begin{tabular}{lccccccc}
\hline
Ref. (year) 
& Band(s) 
& BW (\%) 
& B2B Iso.
& Pol. 
& Scan range 
& Scan loss (dB) 
& Max. AR (dB) \\ 
\hline

\cite{Chang2022NearFieldCircularPolarizer} (2022) 
& K/Ka 
& 20/29.8
& N.G
& Orth. CP$^{a}$ 
& 2-D $(\pm55^\circ/\pm60^\circ)$ 
& 4.4/3.7$^{b,c}$ 
& $<4$ \\

\cite{Kim2022WidebandCPWideARScanning} (2022) 
& 5.8\,GHz 
& 20
& N.A
& DLP/DCP 
& 1-D $(\pm57^\circ)$ 
& $>7^{b}$ 
& $<2$ \\

\cite{Hao2023KKaWideCoverageLEO} (2023) 
& K/Ka 
& 18/20
& 40
& Orth. CP$^{a}$ 
& 2-D $(\pm60^\circ/\pm60^\circ)$ 
& $>5/>7^{b}$ 
& $<5$ \\

\cite{Liu2024LowProfileDualBandCP} (2024) 
& 13/18\,GHz 
& 14.3/10.9
& N.G
& Orth. CP$^{a}$ 
& 2-D $(\pm26^\circ/\pm30^\circ)$ 
& 3$^{d}$ 
& 3/4.6$^{e}$ \\

\cite{Zhang2024WideAngleScanningCP} (2024) 
& L/S 
& 10.5/2.8
& 12/14
& Orth. CP$^{a}$ 
& 1-D $(\pm65^\circ/\pm60^\circ)$ 
& 2.2/2.5 
& 4.7/4.9 \\

\cite{Liang2025KKaARSFSharedAperture} (2025) 
& K/Ka 
& 9.9/2.3
& 27/23
& CP$^{f}$ 
& 2-D $(\pm55^\circ/\pm60^\circ)$ 
& $\leq5/\leq5$ 
& $\leq5/\leq4$ \\

\textbf{This (2026)}
& \textbf{S/X}
& \textbf{21.1/8.1}
& \textbf{34/17}
& \textbf{DLP/DCP} 
& \textbf{2-D ($\mathbf{\pm50^\circ/\pm55^\circ}$)}
& \textbf{4.3/4.1} 
& $\mathbf{\leq 0.1/\leq 0.1}$ \\ 

\hline
\end{tabular}
}

\vspace{1mm}
\begin{minipage}{0.98\textwidth}
\scriptsize
$^{a}$Orth. CP indicates orthogonal CP states assigned to different bands rather than simultaneous co-polarized operation for LHCP or RHCP in the same band. DLP: dual linear polarization. DCP: dual circular polarization. B2B Iso: band-to-band isolation. Scan loss given for worst case.
$^{b}$Estimated from published radiation-pattern or gain plots since numerical scan loss was not explicitly reported.
$^{c}$Simulated scan-loss value.
$^{d}$Reported as gain variation.
$^{e}$Reported within the 3-dB beamwidth.
$^{f}$Dual-CP model was simulated, whereas the measured prototype supported RHCP. 
\end{minipage}
\end{table*}

Table~\ref{tab:AR_comparison} compares the proposed SAPAA with representative recent CP phased-array prototypes. The entries are ordered by publication year, and the footnotes identify scan-loss values estimated from plots, simulated values, and cases where orthogonal CP states are assigned to different frequency bands. The comparison shows that previously reported wide-angle CP phased arrays typically exhibit worst-case scanned AR values of several decibels, often together with scan-loss values comparable to or higher than those of the proposed array. In contrast, the proposed SAPAA demonstrates below-0.1-dB AR in both the S- and X-band receiving bands while maintaining shared-aperture operation, LP/RHCP/LHCP receive-polarization reconfigurability, and experimentally validated wide-angle scanning. This performance is enabled by the combined embedded dual-polarized element response, shared-aperture integration, and calibrated receive-array polarization synthesis.

The X-band total active reflection coefficient and active inter-band isolation were evaluated in simulation, but not experimentally validated for the \(3\times4\) prototype, since obtaining the 68-port S-matrix would require an impractically large number of manual multiport VNA measurements. The virtual-VNA concept presented in \cite{virtualVNA} may provide a practical route for experimental validation of active matching and inter-band isolation in future large shared-aperture arrays.

The far-field measurements were performed in an anechoic chamber using a dual-polarized measurement setup with channel calibration, enabling coherent extraction of the orthogonal field components used in the AR calculation. The RFSoC-based receiver also provides high-resolution complex excitation control, so the amplitude and phase quantization of the synthesized weights is not expected to be the limiting factor. Nevertheless, an AR level of 0.1\,dB is close to the practical limit of polarimetric far-field measurements, since it corresponds to very small residual amplitude and phase imbalances between the two orthogonal components. Therefore, measured values near or below 0.1\,dB are interpreted as demonstrating an uncertainty-limited near-ideal CP state and very high polarization purity, rather than as an absolute metrological claim of sub-0.1\,dB accuracy.

The remaining simulation--measurement differences are consistent with practical prototype and measurement tolerances. The fabricated antenna holder did not include the intended M2 threads, and nylon nuts were therefore added to secure the structure. This caused the screw ends to protrude slightly above the designed height and introduced additional dielectric material above the dipole level. These effects are relevant at X-band because the screws intentionally act as scan-dependent inductively reactive loading pins, and the diagonal steering planes are aligned with the loading-pin directions. The RFSoC measurements further validate simultaneous multi-channel active reception, although the X-band case required an external down-conversion chain and is therefore more affected by mixer and IF calibration uncertainty than the passive chamber measurements. Despite these practical limitations, the passive and active measurements show consistent scan-gain trends and confirm that calibrated receive coefficients recover ultra-low AR in the proposed shared aperture. The wide scan range also supports the intended frustum implementation, since fewer planar SAPAA faces are required to cover the target ground-station field of view. 

\section{Conclusions}\label{ch:conclusion}

A modular S-/X-band SAPAA for polarization-reconfigurable SATCOM ground-station reception was presented. The shared-aperture unit cell combines separately optimized S- and X-band dual-polarized radiators with calibrated complex receive coefficients for LP, RHCP, and LHCP synthesis. Theoretical estimates for scan-dependent loading and inter-band coupling were used to reduce the design burden before full shared-aperture verification. Simulations, passive chamber measurements, and RFSoC-based active measurements confirmed below-0.1-dB AR within the target bands and scan range, together with acceptable scan gain, active matching, and inter-band isolation. The results demonstrate that the proposed scalable SAPAA is a promising building block for agile multi-mission SATCOM ground terminals. Future work for the SAPAA includes experimental validation of antenna profile reduction, additional inter-band isolation enhancement methods, band extension, and investigation of measurement techniques using the virtual-VNA concept.

\section*{Acknowledgment}

The authors acknowledge the use of the MIDAS infrastructure of Aalto University School of Electrical Engineering.

\bibliographystyle{IEEEtran}
\bibliography{IEEEabrv,References}
\end{document}